\documentclass[10pt,a4paper,twocolumn]{article}
\usepackage{f1000_styles}
\usepackage{comment}
\usepackage{natbib}

\usepackage[symbol]{footmisc}

\usepackage{csvsimple}
\usepackage{datatool}

\begin{document}
\pagestyle{fancy}

\title{Atacama Large Aperture Submillimeter Telescope \mbox{(AtLAST)} Science: Our Galaxy} 
\titlenote{}
\author[1]{Pamela Klaassen}
\author[2]{Alessio Traficante}
\author[3]{Maria T. Beltr\'an}
\author[4]{Kate Pattle}%
\author[1]{Mark Booth}%
\author[5]{Joshua B. Lovell}%
\author[6]{Jonathan P. Marshall}%
\author[7,8]{Alvaro Hacar}
\author[9]{Brandt A. L. Gaches}
\author[10]{Caroline Bot}
\author[11]{Nicolas Peretto}
\author[12]{Thomas Stanke}
\author[13]{Doris Arzoumanian}
\author[11]{Ana Duarte Cabral}
\author[14,15]{Gaspard Duch\^ene}
\author[16]{David J. Eden}
\author[17]{Antonio Hales}
\author[18]{Jens Kauffmann}%
\author[19,20]{Patricia Luppe}
\author[21]{Sebastian Marino}
\author[12]{Elena Redaelli}
\author[22]{Andrew J. Rigby}
\author[23]{\'Alvaro S\'anchez-Monge}
\author[2]{Eugenio Schisano} %
\author[24]{Dmitry A. Semenov}%
\author[12]{Silvia Spezzano}
\author[22]{Mark A. Thompson}
\author[25]{Friedrich Wyrowski}

\author[26]{Claudia Cicone}
\author[27]{Tony Mroczkowski}
\author[28]{Martin A. Cordiner}
\author[29,30,31,32]{Luca Di Mascolo}
\author[33,34]{Doug Johnstone}
\author[27]{Eelco van Kampen}
\author[35,36]{Minju M. Lee}
\author[12,37]{Daizhong Liu}
\author[38]{Thomas J. Maccarone}
\author[4,25]{Am\'elie Saintonge}
\author[11]{Matthew Smith}
\author[39]{Alexander E. Thelen}
\author[26,40]{Sven Wedemeyer}

\affil[1]{UK Astronomy Technology Centre, Royal Observatory Edinburgh, Blackford Hill, Edinburgh EH9 3HJ, UK}
\affil[2]{IAPS-INAF, Via Fosso del Cavaliere, 100, I-00133 Rome, Italy}
\affil[3]{INAF -- Osservatorio Astrofisico di Arcetri, Largo Enrico Fermi 5, Firenze 50125, Italy}
\affil[4]{Department of Physics and Astronomy, University College London, Gower Street, London WC1E 6BT, UK}
\affil[5]{Center for Astrophysics, Harvard \& Smithsonian, 60 Garden Street, Cambridge, MA 02138-1516, USA}
\affil[6]{Academia Sinica Institute of Astronomy and Astrophysics, 11F of AS/NTU Astronomy-Mathematics Building\\No.1, Sect. 4, Roosevelt Rd, Taipei 10617, Taiwan}
\affil[7]{Department of Astrophysics, University of Vienna, Türkenschanzstrasse 17, 1180, Vienna, Austria}
\affil[8]{Leiden Observatory, Leiden University, PO Box 9513, 2300-RA Leiden, The Netherlands}
\affil[9]{Department of Space, Earth and Environment, Chalmers University of Technology, Gothenburg SE-412 96, Sweden}
\affil[10]{Observatoire Astronomique de Strasbourg, Universit\'{e} de Strasbourg, UMR 7550, 11 rue de l'Universit\'{e}, F-67000 Strasbourg, France}
\affil[11]{School of Physics \& Astronomy, Cardiff University, The Parade, Cardiff CF24 3AA, UK}
\affil[12]{Max Planck Institute for Extraterrestrial Physics, Gie{\ss}enbachstra{\ss}e 1, 85748 Garching bei M\"unchen, Germany}
\affil[13]{National Astronomical Observatory of Japan, Osawa 2-21-1, Mitaka, Tokyo 181-8588, Japan}
\affil[14]{Univ. Grenoble Alpes, CNRS, IPAG, F-38000 Grenoble, France}
\affil[15]{Department of Astronomy, University of California, Berkeley CA, 94720, USA}
\affil[16]{Armagh Observatory and Planetarium, College Hill, Armagh, BT61 9DB}
\affil[17]{National Radio Astronomy Observatory, 520 Edgemont Road, Charlottesville, VA 22903-2475, United States of America}
\affil[18]{Haystack Observatory, Massachusetts Institute of Technology, Westford, MA 01886, USA}
\affil[19]{School of Physics, Trinity College Dublin, College Green, Dublin 2, Ireland}
\affil[20]{Astrophysikalisches Institut und Universitäts-Sternwarte, Friedrich-Schiller-Universität Jena, Schillergäßchen 2-3, 07745 Jena, Germany}
\affil[21]{Department of Physics and Astronomy, University of Exeter, Stocker Road, Exeter EX4 4QL, UK}
\affil[22]{School of Physics and Astronomy, University of Leeds, Woodhouse Lane, Leeds LS2 9JT, UK}
\affil[23]{Institut de Ci\`encies de l'Espai (ICE, CSIC), Carrer de Can Magrans s/n, E-08193, Bellaterra, Barcelona, Spain}
\affil[24]{Max-Planck-Institut f\"{u}r Astronomie, K\"{o}nigstuhl 17, D-69117 Heidelberg, Germany}
\affil[25]{Max-Planck-Institut für Radioastronomie, Auf dem Hügel 69, 53121 Bonn, Germany}
\affil[26]{Institute of Theoretical Astrophysics, University of Oslo, PO Box 1029, Blindern 0315, Oslo, Norway}
\affil[27]{European Southern Observatory (ESO), Karl-Schwarzschild-Strasse 2, Garching 85748, Germany}
\affil[28]{Astrochemistry Laboratory, Code 691, NASA Goddard Space Flight Center, Greenbelt, MD 20771, USA.}
\affil[29]{Laboratoire Lagrange, Université Côte d'Azur, Observatoire de la Côte d'Azur, CNRS, Blvd de l'Observatoire, CS 34229, 06304 Nice cedex 4, France}
\affil[30]{Astronomy Unit, Department of Physics, University of Trieste, via Tiepolo 11, Trieste 34131, Italy}
\affil[31]{INAF -- Osservatorio Astronomico di Trieste, via Tiepolo 11, Trieste 34131, Italy}
\affil[32]{IFPU -- Institute for Fundamental Physics of the Universe, Via Beirut 2, 34014 Trieste, Italy}
\affil[33]{NRC Herzberg Astronomy and Astrophysics, 5071 West Saanich Rd, Victoria, BC, V9E 2E7, Canada}
\affil[34]{Department of Physics and Astronomy, University of Victoria, Victoria, BC, V8P 5C2, Canada}
\affil[35]{Cosmic Dawn Center (DAWN), Copenhagen, Denmark}
\affil[36]{DTU-Space, Technical University of Denmark, Elektrovej 327, DK2800 Kgs. Lyngby, Denmark}
\affil[37]{Purple Mountain Observatory, Chinese Academy of Sciences, 10 Yuanhua Road, Nanjing 210023, China}
\affil[38]{Department of Physics \& Astronomy, Texas Tech University, Box 41051, Lubbock TX, 79409-1051, USA }
\affil[39]{Division of Geological and Planetary Sciences, California Institute of Technology, Pasadena, CA 91125, USA.}
\affil[40]{Rosseland Centre for Solar Physics, University of Oslo, Postboks 1029 Blindern, N-0315 Oslo, Norway}
\maketitle
\thispagestyle{fancy}

\clearpage
\begin{abstract}

As we learn more about the multi-scale interstellar medium (ISM) of our Galaxy, we develop a greater understanding for the complex relationships between the large-scale diffuse gas and dust in Giant Molecular Clouds (GMCs), how it moves, how it is affected by the nearby massive stars, and which portions of those GMCs eventually collapse into star forming regions. The complex interactions of those gas, dust and stellar populations form what has come to be known as the ecology of our Galaxy. Because we are deeply embedded in the plane of our Galaxy, it takes up a significant fraction of the sky, with complex dust lanes scattered throughout the optically recognizable bands of the Milky Way. These bands become bright at (sub-)millimetre wavelengths, where we can study dust thermal emission and the chemical and kinematic signatures of the gas. To properly study such large-scale environments, requires deep, large area surveys that are not possible with current facilities. Moreover, where stars form, so too do planetary systems, growing from the dust and gas in circumstellar discs, to planets and planetesimal belts. Understanding the evolution of these belts requires deep imaging capable of studying belts around young stellar objects to Kuiper belt analogues around the nearest stars. Here we present a plan for observing the Galactic Plane and circumstellar environments to quantify the physical structure, the magnetic fields, the dynamics, chemistry, star formation, and planetary system evolution of the galaxy in which we live with AtLAST; a concept for a new, 50m single-dish sub-mm telescope with a large field of view which is the only type of facility that will allow us to observe our Galaxy deeply and widely enough to make a leap forward in our understanding of our local ecology.

\end{abstract}

\section*{\color{OREblue}Keywords}

Astronomical instrumentation, methods and techniques;
Telescopes;
The Galaxy;
Galaxy: solar neighborhood;
Submillimeter: planetary systems;
Submillimeter: ISM;
Magnetic fields;
Surveys

\pagestyle{fancy}

\clearpage
\section*{Plain language summary}

There are many individual components contributing to the overall evolution of our Galaxy, the Milky Way. Through understanding the physics and chemistry of the Galaxy around us, we better understand our origins, our environment, and where we're going. Here we outline a number of observational surveys of our Galaxy that would produce a step change in our understanding of the evolution of the Galaxy around us, both as a template for others, and as the only way of understanding our place in the larger Universe.  We present surveys of the Galactic Plane focusing on the dust and magnetic fields, chemistry, and dynamics of the gas. We then suggest targeted observations of local stars and star forming regions to uncover the origins of stars, planets and how those planetary systems evolve over the course of their lives, helping to put our Sun and Solar System in context. These types of observations require simultaneously sensitive, long wavelength (between 0.3 and 3 millimetre) observations as well as a large coverage of the sky, and cannot be done with current observatories operating at these wavelengths. Future leaps in understanding these systems will require a new telescope; a large telescope at a good observing location with a large field of view. This telescope, the Atacama Large Sub-mm Telescope (AtLAST)\footnote{\url{http://atlast-telescope.org/}} is being developed, and here we are presenting the science cases for this new telescope from the point of view of our Galaxy. Together, these studies will revolutionise our understanding of the history and evolution of our Galaxy and bring us yet another step closer to understanding our place in, and the evolution of, our Universe.

\section{Introduction}

Galactic ecology is the emerging field of understanding our Galaxy as an unfolding system with many interconnected components that, together, shape its evolution.  As with an ecosystem on Earth, there are many individual components contributing to the overall evolution of the system, our Galaxy. Through understanding the physics and chemistry of the galaxy around us, we better understand our origins, our environment, and where we're going.  Integral to this is quantifying how gas moves through the stages from diffuse clouds to dense filaments and eventually star forming cores and stellar systems, which requires observations in the mm and (sub-)mm ($0.3 - 3$~mm) wavelength ranges.

\begin{figure*}[hbt]
    \centering
    \includegraphics[width=1\linewidth]{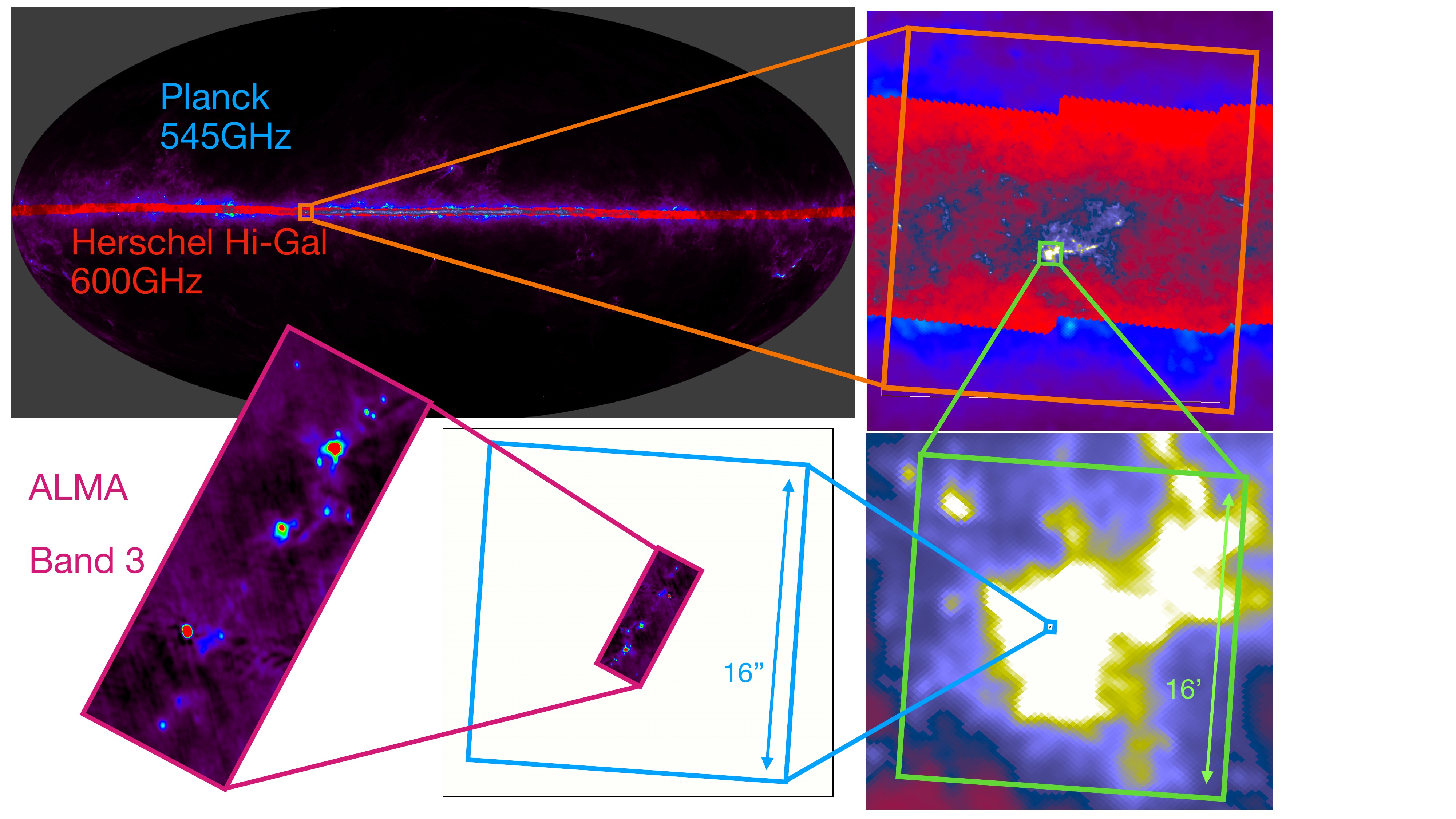}
    \caption{Gradual zooming view on the widely different spatial scales from the full sky dust continuum emission as seen by Planck at low resolution up to a typical ALMA observation at high resolution. The top left image shows the Planck all-sky observations from the High Frequency Instrument at 545GHz \citep{Planck2020HFI} for the whole sky, together with the Herschel Hi-Gal survey \citep{Molinari2010} observations at 600GHz cover a large part of the galactic plane as an overlay. From there to the bottom left ALMA images are individual zoom-in views on the Hi-GAL survey, up to the W51 north region where an ALMA observation exists. This ALMA image is taken as an example of a typical observation and corresponds to the band 3 image of W51 north from \citet{Tang2022}. The two last zoom-in regions represented by the green and blue box correspond to 16' and 16" wide regions respectively. }
    \label{fig:planck-to-alma}
\end{figure*}

Our current understanding of the ecology of our Galaxy is limited  by technology (see, for example, Figure \ref{fig:planck-to-alma}). With (sub-)mm interferometers like ALMA we can delve deep into the details of known targets of interest on size scales of up to a few square arcminutes. With Far IR space telescopes like \textit{Herschel} and \textit{Planck} we can probe large scale dust emission at low resolution on size scales of hundreds of sq. deg (i.e. the whole plane).  With current (sub-)mm single dish facilities we can get some of those intermediate scales, but with limited resolution, fields of view and sensitivity, we are restricted to areas of up to a few sq. deg.

To understand our Galaxy as a whole requires more than these heterogeneous sets of observations which probe very different size scales, and thus physics and chemistry, can provide.  We need dedicated, statistically significant surveys of the Galaxy to properly put these hand-picked regions into context.

Being our own galaxy, the Milky Way is studied, not only as the environment in which our Sun and Solar System have come to exist, but as the nearest example of a dynamic and evolving galaxy.  It is the galaxy for which we will get the highest resolution (including the resolved observation of discs and planets) and hence the most detailed view on processes like star formation, astrochemistry, the lifecycle of stars, the thermodynamical evolution of the Interstellar Medium (ISM), as well as the place where discs and planets can be best detected and resolved. Because of this level of detail, the Milky Way is also a reference point for nearby galaxy studies (See also companion article by Lee et al.). Nearby galaxies provide an outsider view point as well as enough resolution to observe individual environments (molecular clouds or clumps, HII regions, clusters, etc.) in different conditions (metallicity, galaxy mass, environment, etc.), but are studied with the galactic context as a reference. Because processes like star formation are multi-scale and in order to make the comparison between very resolved processes observed in our Galaxy with the ones observed in nearby galaxies, large Galactic Plane surveys have proven to be key to understanding both resolved structures (molecular clumps, cores, filaments, discs, \ldots) and to get a ``global'' view by sampling a wide range of spatial scales. For example, the GLIMPSE, Hi-GAL and the ATLASGAL surveys of the galactic plane in the infrared and sub-millimeter with Spitzer, Herschel and APEX, respectively, have transformed our view of gas, dust, star formation and young stellar objects within our galaxy and provided legacy references for comparison to different environmental conditions.

Small aperture, wide area observatories are able to give a low resolution view of our Galaxy, and interferometers like the Atacama Large Millimeter Array (ALMA) are able to probe the finer points in exquisite detail.  What is missing is the intermediate (sub-pc to few-pc) scales that join these two regimes of observations.  For this, a new, large, single-dish (sub-)mm facility is required.  With the Atacama Large Aperture Sub-mm Telescope  (AtLAST) \citep{2020Klaassen,2022Ramasawmy,2023Mroczkowski,Mroczkowski2024}, a concept for a 50-m class (sub-)mm single dish telescope with a wide field of view and throughput, we enter a new regime in understanding the interactions of the gas and dust throughout the plane of our Galaxy, and thus how it evolves as an ever-changing ecosystem.

Despite its importance, the large angular scales and wide frequency coverage required to efficiently survey our Galaxy have traditionally limited its characterization to specific wavelengths. Only low-resolution (i.e., few 10-100s of arcseconds) continuum surveys, usually carried out by space observatories such as Planck, have been able to achieve all-sky maps of the Milky Way \citep{Planck2015, Planck2020}. Herschel observations provided higher resolution observations in the FIR (i.e., down to 5$''$), although covering smaller areas typically around the Galactic Plane \citep[e.g.][]{Molinari2010} or selected nearby regions (i.e., Orion, Taurus). In terms of line emission, significant efforts have been made to produce HI maps at centimeter wavelengths (such as LAB, \citet{Kalberla2005} or HI4PI, \citet{HI4PI2016} surveys) tracing the atomic gas in our Galaxy. Higher spatial and spectral resolution observations to study the atomic contribution to the Cold Neutral Medium (CNM) and Warm Neutral Medium (WNM) at low-frequency in the galactic plane are being carried out with projects like GASKAP \citep{Dickey2013}, paving the way to even higher resolutions in the future by interferometers such as SKA and ngVLA. In contrast, molecular line observations are broadly either wide-area and low resolution (i.e., 10s of arcsec to arcminute resolution) surveys \citep{Dame2001, Schuller17,Schuller2021} or small targeted high resolution observations with facilities such as ALMA.

The above limitations are particularly acute at \mbox{(sub-)mm} wavelengths. Sensitivity has traditionally restricted the observations of small, single-dish telescopes to dedicated Galactic plane surveys of bright emission lines \citep{Dame2001, Schuller17, Schuller2021} or peculiar targets (e.g., the Central Molecular Zone, CMZ) at low resolution. Large interferometers such as ALMA, on the other hand, have increased our ability to detect both line and continuum emission at high resolutions, but are instead restricted to small Fields-of-View (FoV). The Fred Young Submillimeter telescope (FYST, formerly CCAT-p) is expected to cover significant fractions of the celestial sphere in the highest frequency (sub-)mm regime, but with a 6m primary mirror, will be limited in sensitivity and resolution. These studies are intended to provide vital information on the physical and chemical evolution of both molecular and dust components in the CNM, critical for understanding the evolution of our Galaxy. However, these large area surveys come at the cost of much lower spatial resolution. %
Compared to recent breakthroughs provided by high-resolution studies, the characterization of the large-scale properties of the cold ISM at (sub-)mm wavelengths in our Galaxy remain largely unexplored.

Here we outline a number of Galactic science cases that would produce a step change in our understanding of the evolution of the Galaxy around us, both as a template for others, and as the only way of understanding our place in the larger Universe.  We present surveys of the Galactic Plane focusing on the (dust) continuum and magnetic fields, chemistry, and dynamics of the gas. We then present a concept for targeted observations of local stars and star forming regions to uncover the statistics around the origins of stars and  planets and around how planetary systems evolve over the course of a stars' life to help put our Sun (see companion article by Wedemeyer et al.) and Solar System (see companion article by Cordiner et al.) in context.

Through the Galactic polarisation project, we will be able to map the polarisation signals produced by the magnetic fields threading our Galaxy along with the dust continuum emitted at the same wavelengths. By linking from protostellar core scales to that of the Galaxy as a whole, we will finally be able to quantify the relative importance, origin and evolution of the magnetic fields in these regions.  Mapping 540~sq.\ deg in multiple bands (230, 345 and 460~GHz) using wide bandwidth (i.e., 32~GHz) receivers down to polarisation fractions of 2\% gets to the same polarisation fractions achieved by Planck on the even larger size scales, linking our Galaxy to its environment. To achieve these unprecedented polarisation sensitivities requires deep integrations that will simultaneously probe the dust properties of the regions being studied.  At these wavelengths, in the Galactic Plane, the thermal dust component becomes the brightest emitter, which allows us to probe its density, temperature and physical structure \textit{all at once}.

In the chemistry program, we add another layer to our understanding of the interplay between gravity, turbulence and magnetic fields derived above by quantifying the heating and cooling in the Galaxy, which of the simplest `complex organic molecules' (COMs) are available for planet forming discs, and how all of these tie into the lifecycle of gas in our Galaxy.  This program aims to reach a sensitivity of 0.3~K per 0.1~km~s$^{-1}$ channel in the 230, 345, 460 and 690~GHz windows across the entire galactic plane, and then delving deeper (0.03~K in 0.01~km~s$^{-1}$ channels) in regions identified in the shallower full plane survey. This will not only allow us to probe dense gas tracers, and to create a CO spectral line energy distribution (SLED) across the entire galaxy, but also detect complex organic molecules  within a few kpc. Thanks to the combination of the high sensitivity and high resolution that AtLAST offers, in fact, it will be able to detect the most abundant COMs (methanol and formaldehyde) and also some of the less abundant, more complex species (e.g. methyl formate, dimethyl ether, ethanol, acetaldehyde). These can be locally enhanced, for instance, by the heating due to protostellar feedback in hot cores and corinos.

With the Galactic dynamics survey, we will be able to trace both the physical properties and the gas dynamics of a volume limited (10~kpc) sample of star forming regions in our Galaxy.  This will be the first comprehensive and statistically significant study of the gas dynamics in star-forming regions in different Galactic environments. It aims to reach a sensitivity of 0.01~K per 0.05~km~s$^{-1}$ channel, which is enough to detect key lines (e.g., CO and N$_2$H$^+$ which respectively trace diffuse and dense cold gas in the 3~mm band) across the entire Galactic plane. The 50~m dish both enables those sensitivity limits to be reached, as well as providing the ability to see sub-parsec scale structures (i.e., star forming cores) up to 10~kpc away.

These observations will bring about a new era in our understanding of the history and evolution of our galaxy, because they would, for the first time, unlock large area surveys of the closest galaxy we can study.

The unprecedented sensitivity \textit{and field of view} of AtLAST will also enable deep studies of small areas, such as star forming regions, where we will be able to conduct a complete census of young stars, tracing the evolution from protoplanetary to debris discs.

Closer to home, and often outside the plane of the galaxy, AtLAST will provide the ability to study the very nearest planetary systems. It will be able to detect the second generation dust created by belts of remnant planetesimals. So far, most detected debris discs are orders of magnitude more massive than our own Kuiper belt, which lost much of its mass during a phase of dynamical rearrangement of the planets. With AtLAST, we will be able to detect discs as faint as our own Kuiper belt around other stars, allowing us to place our Solar System into context. The resolution at the shortest wavelengths will make it possible to resolve structures in the disc down to scales of a few au, whilst retaining sensitivity to large-scale structure that is not possible with interferometers. These structures are often our best tool for understanding the outer reaches of planetary systems where only the brightest of planets are currently detectable.

Overall, the ambition of these projects far exceeds the capabilities of current and near-term facilities. No currently operating, or under construction facility can address these questions about our origins and place in the dynamic Universe.  A 50~m class (sub-)mm single dish telescope with a large field of view populated with mega-pixel continuum (polarisation) cameras capable of simultaneous multi-chroic observations and kilo-pixel high-spectral resolution (e.g., 0.1~km~s$^{-1}$ or higher) instruments, all of which have bandwidth comparable to those expected in the ALMA Wideband Sensitivity Upgrade, will enable us to answer these questions, and shape the next questions to ask after that.

AtLAST is a 50~m-class single-dish, sub-millimeter telescope to be installed at the Chajnantor Plateau. Covering a frequency range between 100 $\sim$900~GHz, its large collecting area will be paired with a new generation of state-of-the-art, mega-pixel bolometers and kilo-pixel heterodyne cameras, and up to 2~degree Field of View providing AtLAST with unprecedented throughput. Compared to current and planned single-dish telescopes, AtLAST is expected to improve the mapping speed and sensitivity of previous continuum and molecular surveys by more than two order of magnitudes, getting well below the confusion limits of its predecessors at these wavelengths. These characteristics will enable for the first time Galaxy-wide studies on molecular lines and dust emission, as well as polarisation. With no other similar facility in the southern hemisphere, AtLAST will cover a unique niche in terms of combined spatial coverage, sensitivity, and resolution for both galactic and extragalactic studies. %

 In this paper, we focus on the current state of the art in studies of our Galaxy, where we expect we could make the greatest gains given new technology, and what that facility/instrumentation suite would need to have in order to make those gains.  We separate the paper into scientific themes below, focusing on Galactic dust and magnetic fields (Section \ref{sec:Pol}) and  dynamics (Section \ref{sec:Dyn}), followed by sections outlining cases for studying detailed chemistry (Section \ref{sec:Eco}) and stellar and star forming regions (Section \ref{sec:discs}).  We then describe the technical requirements this imposes on a new facility (Section \ref{sec:telescope}), and then summarise these final requirements in a matrix based on the individual science cases.

\section{The Magnetic Field and Gas Density Structure of Our Galaxy}
\label{sec:Pol}

\newcommand{\Pol}[1]{\DTLfetch{projects}{Parameters}{#1}{Galactic_Polarisation}}

\subsection{Scientific Rationale}

The interstellar medium (ISM) of the Milky Way is a complex, non-linear system in which the interplay between gravity, turbulence, magnetic fields and feedback takes place over many orders of magnitude in size scale.  As such, it has become increasingly apparent that the ISM must be considered as a whole, if we are to understand how the fundamental properties of star formation, such as star formation rate (SFR) and efficiency (SFE), and the Initial Mass Function (IMF) are set.

Dust emission provides an optically thin tracer of column density \citep{Hildebrand1983}, and when observed in polarised light, also allows us to map the plane-of-sky interstellar magnetic field \citep{hiltner1949,hall1949,davisgreenstein1951}.  With AtLAST it will be possible for the first time to undertake an unbiased multi-wavelength polarised continuum survey of the Galactic Plane at high enough spatial resolution to resolve individual star-forming cores.  This will give us unprecedented insight into the dynamics, energetic balance and magnetic field properties of the interstellar medium, and so into the physics of star formation.  Moreover, as the ISM is typically only a few percent polarised \citep{planck2015_intXIX}, by performing this magnetic field mapping, we will also gain a map of the dust column density of the Galactic Plane, measured with unprecedented sensitivity and angular resolution. 

Submillimetre-wavelength observations are crucial to this undertaking, as at these wavelengths, continuum polarisation arises from dust grains aligned with respect to the magnetic field (e.g. \citealt{andersson2015}).  At wavelengths $> 1$\,mm, continuum emission increasingly arises from thermal free-free emission (no net polarisation; \citealt{rybicki1979}) or synchrotron radiation (up to $\sim 40$\% polarisation; e.g. \citealt{dubner2015}).  By observing at 230\,GHz (1.3\,mm), 350\,GHz (0.86\,mm) and 460\,GHz (0.65\,mm), we will be able to accurately characterize the polarised and unpolarised spectral energy distributions (SEDs) of dust in the Galactic plane.  The polarised SED will allow us to measure ISM magnetic field morphology and infer magnetic field strength, as well as to uncover key information on dust composition and physics.  Meanwhile the unpolarised SED will allow us to measure dust column density and, in combination with shorter-wavelength \textit{Herschel} observations, dust temperature and spectral index.

To achieve our key science goal of mapping magnetic fields from cloud scales to individual star-forming cores across the Galaxy, we require a total-power instrument with resolution of a few arcseconds.  Interferometers are unsuitable for such work because their fields of view are too small for them to serve as survey instruments on the scale of the Galactic plane, and because we must observe the full range of spatial scales if we are to trace magnetic field from Galactic to core scales in the ISM.  However, this survey will provide a wealth of targets for interferometric follow-up.  Moreover, since the resolution of AtLAST will be comparable to the primary beam size of submillimetre interferometers such as ALMA, there will be the potential to synthesise AtLAST data with interferometric follow-up observations.  Similarly, although AtLAST continuum data will be spatially filtered on scales comparable to the camera FoV (i.e. $\sim 1^{\circ}$), we will be able to synthesize our data with lower-resolution \textit{Planck} all-sky measurements to restore these larger spatial scales, thereby providing continuous mapping of the ISM from Galactic to au scales.

We next discuss some of the detailed scientific questions which this survey would address.

\begin{figure*}
    \centering
    \includegraphics[width=0.7\textwidth]{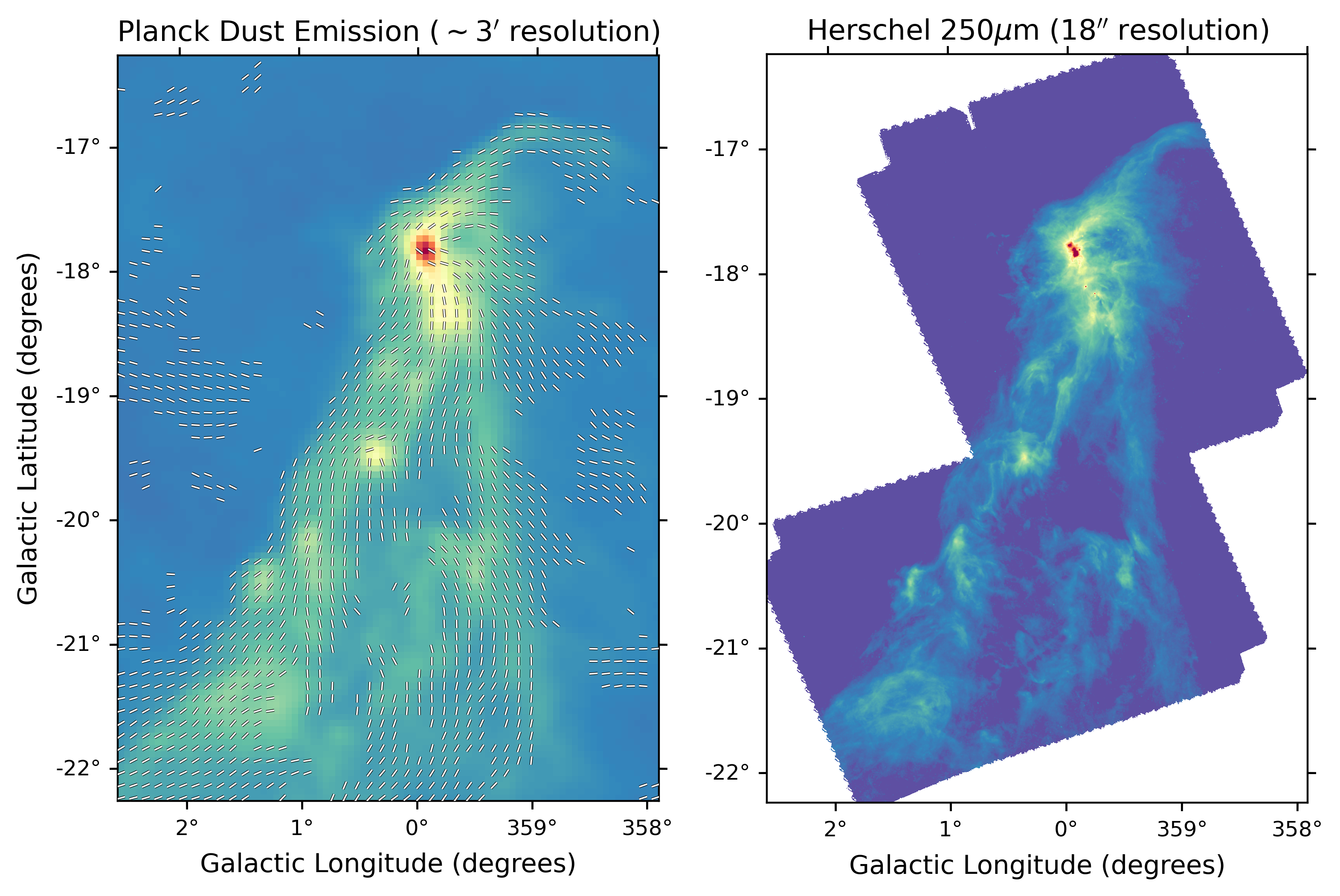}
    \caption{Comparison of Planck 353\,GHz (\citealt{planck2015_intXIX}; left) and Herschel 1.2\,THz (250\,$\mu$m; \citealt{bresnahan2018}; right) mapping of the Corona Australis molecular cloud.  While the Planck data shows the overall shape of the magnetic field in the cloud, it cannot resolve individual sites of star formation.  With AtLAST, we will map magnetic fields across the entire Galactic plane with a resolution at 345\,GHz that is four times better than that shown in the right-hand panel, thereby resolving individual star-forming cores within molecular clouds.}
    \label{fig:cra}
\end{figure*}

\subsubsection{Interstellar magnetic fields}

Magnetic fields are important in ISM substructure formation and play a dynamically crucial role in a wide range of astrophysical processes, from Galactic to cosmological scales, including the evolution of the ISM, star formation, shocks, and cosmic ray acceleration and transport \citep{han2017}. Therefore, observations of the magnetic field have important implications for understanding astrophysical phenomena in a wide range of environments. However, despite their importance, the origin and evolution of magnetic fields in the Universe are not properly understood (e.g., \citealt{rees2005}).  Despite decades of research, no consensus regarding the role of magnetic fields in star formation has been reached yet (e.g. \citealt{pattle2023}, and refs. therein).  While some theories propose that the star-formation process is magnetically regulated (\citealt{shu1999}; \citealt{mouschovias1999}) and that magnetic fields can significantly influence each stage of star formation, affecting fragmentation and cloud collapse, influencing accretion and driving outflows 
(\citealt{tan2013}; \citealt{tassis2014}; \citealt{susa2015}; \citealt{klassen2017}; \citealt{matsushita2018}), others argue that the process is driven by turbulent flows, especially in the high-mass regime (\citealt{maclow2004}; \citealt{vazquezsemadeni2011}; \citealt{wang2014}). Moreover, the link between magnetic fields in molecular clouds and galactic-scale magnetic fields \citep[e.g.][]{lopezrodriguez2021,borlaff2023}, and particularly the scale at which the role of magnetic fields in counteracting the self-gravity of clouds becomes important, remains uncertain.  To solve this long-standing debate it is necessary to accurately characterise the properties of magnetic fields -- morphology, strength and dynamic importance -- in a statistical way at a Galactic scale, which can help not only to better understand its role in crucial processes such as star formation, but also to gain insights into their origin and evolution.

In the ISM, the presence of magnetic fields is usually revealed by dust polarisation observations, which are an effective and efficient means by which to measure the plane-of-sky direction of the magnetic field \citep{davisgreenstein1951}, and to assess the relative magnitudes of the mean and turbulent components of the field \citep{hildebrand2009}. Magnetic fields can be traced through submillimetre-wavelength observations of linearly-polarised dust continuum emission resulting from dust grains aligned with their minor axes parallel to the local magnetic field direction as a result of radiative torques (RATs) (\citealt{lazarian2007}; \citealt{andersson2015}). %

The polarised emission from thermal dust provides important information on the structure and morphology of the magnetic field, revealing the orientation of the field averaged along the line of sight and projected on the plane of the sky (see, e.g., \citealt{hensley2021}; \citealt{planck2015_intXXII}).
The strength and dynamic importance of the magnetic field can be estimated using techniques such as the Davis-Chandrasekhar-Fermi method (\citealt{davis1951}; \citealt{chandrasekhar1953}) and its recent structure-function based modifications (\citealt{hildebrand2009}; \citealt{houde2009}; \citealt{chen2022}), the polarisation-intensity gradient \citep{koch2012a,koch2012b,koch2014}, or the histogram of relative orientations (HRO; \citealt{soler2013}). Techniques such as polarisation-intensity gradients and HROs study the relative magnetic field orientation with respect to the orientation of intensity or column density structures, thus allowing derivation of position-dependent field strength estimates based on the observed magnetic field morphology. The turbulent velocity dispersion, and so the strength of the turbulent forces, which are needed to quantify the relative importance and strength of magnetic fields can be obtained from the optically thin molecular line observations described in Section~\ref{sec:Dyn}.

\subsubsection{Unbiased mapping of the column density structure of the Galactic Plane}

Numerical simulations demonstrate that an accurate knowledge of initial conditions is crucial to understanding the star formation process (e.g. \citealt{bonnell2011}; \citealt{krumholz2014}). Mapping the column density structure of the Galactic plane at high resolution with AtLAST will allow us to observe all stages of cloud evolution, and to probe the large diversity of Milky Way environments, for example in terms of Galactic structure, interstellar radiation field, or presence of local H\textsc{ii} regions.  The survey will also allow us to map complete samples of star-forming cores across the Galactic plane, and so to connect the properties of star-forming cores throughout the Milky Way to the conditions in which they formed.

Interstellar filaments are ubiquitous within star-forming regions, as demonstrated by \textit{Herschel} surveys of molecular clouds (e.g. \citealt{arzoumanian2011}; \citealt{andre2014}; \citealt{schisano2014}), and their close associations to cores clearly demonstrate that they are a key element to the star formation process. Research on the physics of filament formation and fragmentation has intensified significantly over the last ten years (e.g. \citealt{hennebelle2013}; \citealt{palmeirim2013}; \citealt{ntormousi2016}; \citealt{clarke2017}; \citealt{inoue2018}; \citealt{lu2018}; \citealt{hacar2023}).  A central debate is over whether or not a universal filament width of $\sim 0.1$ pc may exist (e.g. \citealt{arzoumanian2011}; \citealt{panopoulou2017}; \citealt{arzoumanian2019}; \citealt{priestley2022}), and if it does, how it may be linked to the peak of the IMF at $\sim 0.3$ M$_{\odot}$. 

Filaments appear to also play a key role in the formation of high-mass stars.  Hub-filament systems (HFSs), i.e. networks of converging filaments connecting to a high-density clump (e.g. \citealt{myers2009}), have been identified to host all of the most luminous ($>10^5$ L$_{\odot}$) young stellar objects in the Milky Way (\citealt{kumar2020}). It has been proposed that the converging nature of HFSs is the direct result of the global collapse of their parent clumps, facilitating the formation of massive stars at the bottom of their gravitational potential well (e.g. \citealt{Peretto13}; \citealt{williams2018}; \citealt{rigby2024}). However, very little is known about how HFSs form and evolve.  Determining whether or not filament convergence is present from the start (e.g. \citealt{balfour2015}) or a characteristic that comes about as the clumps evolve over time (e.g. \citealt{Vazquez-Semadeni19}) would settle many of the current debates on the initial conditions of massive star formation (e.g. \citealt{peretto2022}).  However, the angular resolution of available (sub-)millimetre continuum surveys is not high enough to resolve the width of interstellar filaments, except within the nearby, low-mass Gould Belt star-forming clouds.  Existing observations thus provide a very biased view of the Milky Way filament population.  An unbiased survey of the Galactic Plane with a resolution of a few arcseconds is necessary in order to gain a statistical understanding of the role HFSs play in high-mass star formation.

\subsubsection{Dust physics}

Multiwavelength submillimetre dust continuum polarisation observations provide us with information on grain alignment mechanisms, and on the physical properties of dust grains \citep{andersson2015}.  This information is encoded in how both polarisation fraction and polarisation angle change as a function of wavelength, and with local environment.

While over the vast majority of interstellar environments dust grains are preferentially aligned with respect to magnetic fields by radiative torques (``B-RAT'' alignment; \citealt{lazarian2007}), in the densest and most highly-irradiated environments a wide range of competing polarisation mechanisms come into play, including standard B-RAT alignment, ``k-RAT'' alignment in which radiative torques align dust grains along the radiation vector \citep{tazaki2017}, mechanical alignment in which dust grains are aligned by gas flows \citep{hoang2018}, and dust self-scattering \citep{kataoka2015}.  Which of these mechanisms dominate in a given region depends sensitively on grain size, composition and environment (e.g. \citealt{pattle2021}).  Multiwavelength polarisation observations are crucial in order to understand which grain alignment mechanism dominates as a function of grain size in a given region (e.g. \citealt{ko2020}).  With the resolution and multiwavelength polarisation capabilities of AtLAST, we will be able to explore how grain alignment mechanisms, and so grain properties, vary across the Galactic plane.

In regions where B-RATs are known to drive grain alignment (the large majority of the ISM; \citealt{lazarian2007}), multiwavelength polarisation observations still provide us with invaluable insights into dust physics and magnetic field structure.  Where different wavelengths trace different magnetic field geometries, our observations may trace different dust components along the line of sight, with hot and cold dust components tracing different magnetic fields (e.g. \citealt{chuss2019}; \citealt{pattle2021a}).  In such cases, multiwavelength polarisation observations provide valuable insight into the three-dimensional structure of both magnetic fields and density structure.  Moreover, by observing at wavelengths $< 1$ mm and at 1.3 mm simultaneously, we will be able to search for changes in the 1.3 mm emission that are indicative of this wavelength tracing free-free or synchrotron, rather than dust polarisation.  This behaviour is most likely to be seen in shocked material in high-mass star-forming regions (free-free; \citealt{fernandezlopez2021}) or in supernova remnants (synchrotron; \citealt{dubner2015}), and so will provide us with valuable insight into the effects of stellar and supernova feedback on star-forming regions and the magnetic fields within them.

Simultaneously, polarisation fraction measurements provide insight into dust grain properties even where a single (B-RAT) grain alignment mechanism dominates.  Ground-based observations of star-forming regions have suggested a minimum in polarisation fraction at 350µm (e.g. \citealt{vaillancourt2002}), with polarisation fractions rising at longer wavelengths.  However, stratospheric observations of Vela C with the BLASTPOL balloon-borne telescope have found a much flatter polarisation spectrum \citep{gandilo2016}.  The nature of the the polarisation spectrum provides information on grain properties within molecular clouds: a rising spectrum with wavelength requires colder grains to be better-aligned with the magnetic field.  However, statistical comparison of polarisation spectra between star-forming regions has not yet been undertaken, and how dependent polarisation spectra are on environment remains unclear.  This survey, by observing polarisation fractions from 230 to 460 GHz, would allow us to map polarisation spectra across the entire Galactic plane.

\subsubsection{Transients}

The current era of transient astronomy is not restricted to the optical part of the electromagnetic spectrum, but is also undergoing significant advances in the far-infrared and sub-millimetre regimes.  While early models of star formation involved steady accretion rates onto the surface of protostars (e.g. \citealt{larson1969}; \citealt{shu1987}), there is now a plethora of evidence that episodic accretion is a common feature of star formation across all stellar masses \citep[e.g.][]{hartmann2016}. One can reasonably expect the formation of massive stars, which occurs within dense, clustered environments, to involve extremely large accretion bursts as a result of the chaotic and dynamic nature of high-mass star formation. In some cases it is predicted that protostellar luminosity can increase by more than 3 orders of magnitudes, with burst durations of a couple of years (e.g. \citealt{meyer2021}).  The corresponding flux increase is largest in the mid- to far-infrared but is still very significant up to wavelengths of a few hundred microns \citep{fisher2022}.  Observations made by the JCMT Transients Survey \citep{mairs2017} have shown that these accretion bursts are detectable at 850 microns, and frequent enough to be identified even when only individual molecular clouds are monitored. A recurrent, yearly, survey of the Galactic plane at submillimetre wavelengths would therefore reveal a completely new picture of the Milky Way: one that shows perpetual transformation. Such a picture would allow us to constrain the accretion history of young stellar objects across the Galaxy, and so be transformational in our understanding of how young stars acquire mass. More details on how AtLAST can contribute to this understanding are presented in the companion paper by Orlowski-Scherer et al.

\subsection{Current and previous facilities}

\paragraph{Millimetre/submillimetre cameras}

There are a number ground-based of millimetre and submillimetre cameras that are currently operating, or recently operational, including the ground-based SHARC-II 350\,$\mu$m (857\,GHz) camera on the CSO (\citealt{dowell2003}; decommissioned), the LABOCA 870\,$\mu$m (345\,GHz) camera on APEX (\citealt{siringo2009}; decommissioned), the SCUBA-2 850\,$\mu$m (353\,GHz) and 450\,$\mu$m (667\,GHz) camera on the JCMT (\citealt{holland2013}; operational), the 1.15\,mm (260\,GHz)and 2\,mm (150\,GHz) NIKA-2 camera on the IRAM 30\,m telescope (\citealt{adam2018}; operational), and the 1.1\,mm (273\,GHz), 1.4\,mm (214\,GHz) and 2\,mm (150\,GHz) TolTEC camera on the LMT (\citealt{wilson2020}; commissioning).  Stratospheric or space-based submillimetre/far-infrared cameras have also flown over the last fifteen years, notably the space-based SPIRE 250, 350 and 500\,$\mu$m (1200, 857 and 600\,GHz) and PACS 70, 100 and 160\,$\mu$m (4.3, 3.0 and 1.9\,GHz) cameras on \textit{Herschel} \citep{griffin2010,poglitsch2010}, the 300\,$\mu$m -- 11\,mm (850 -- 30\,GHz) \textit{Planck} Observatory \citep{planck2014}, and the stratospheric 53 -- 214$\mu$m (5.7 -- 1.4\,THz) HAWC+ camera on SOFIA \citep{harper2018}.
It should be noted that at wavelengths $< 1$\,mm, observations are currently restricted to relatively low resolutions of $\gtrsim 10^{\prime\prime}$, with the best resolution currently being achieved by SCUBA-2 on the 15\,m-diameter JCMT.  AtLAST will thus produce a step change in resolution and mapping speed for continuum observations at submillimetre wavelengths.

Over the last fifteen years, a number of submillimetre and millimetre continuum surveys of the Galactic plane have been completed using these cameras (e.g. \citealt{schuller2009}; \citealt{Molinari2010}; \citealt{eden2017}; \citealt{rigby2021}).  Arguably the most influential of these have been the APEX ATLASGAL survey at 870\,$\mu$m \citep{schuller2009} and the \textit{Herschel} HIGAL survey at 70 -- 500\,$\mu$m \citep{Molinari2010}. Those two surveys have significantly contributed to our current understanding of the physical properties of star-forming regions across the Galaxy, mostly via the compilation of large and unbiased samples of dusty cold clumps at $\sim 20^{\prime\prime}$ resolution, i.e. 0.5 pc at 5 kpc distance (e.g. \citealt{urquhart2018}; \citealt{Elia21}). Those clumps serve as primary targets for follow up observations at higher angular resolution with interferometers such as ALMA. For instance, two of the ALMA large programmes on star formation, ALMA-IMF \citealt{motte2022} and ALMAGAL (P.I.: S. Molinari; see, e.g. \citealt{jones2023}), have used both ATLASGAL and HIGAL source catalogues to select their target lists so that core populations at a resolution of $\sim 1000$ AU could be mapped out, with the aim of understanding how the IMF of stars is established. However, the ALMA-IMF and ALMAGAL large programmes had to compromise between either ``complete'' mosaicing of a of handful of clouds or single-pointing observations of a few thousand clumps, neither of which is ideal.  AtLAST's ability to perform a full, unbiased continuum survey of the Galactic plane at arcsecond resolution would allow for the first time a true statistical understanding of star-forming cores in the Milky Way.

\paragraph{Millimetre/submillimetre polarimeters}

The last decades have witnessed a rapid development of submillimetre and far-IR polarimetric facilities. These include new capabilities in single-dish telescopes, such as SHARP on the CSO \citep{li2008}, BLAST-Pol \citep{galitzki2014}, PolKa on APEX \citep{wiesemeyer2014}, \textit{Planck} \citep{planck2014}, POL-2 on the JCMT \citep{friberg2016}, HAWC+ on SOFIA \citep{harper2018}, Toltec on the LMT \citep{montana2019}, and NIKA2-Pol on IRAM \citep{ritacco2020}, and interferometers, such as the SMA \citep{marrone2006}, CARMA \citep{hull2015} and ALMA \citep{nagai2016}. These facilities have shed light on the role of magnetic fields in different Galactic environments: molecular clouds, filaments and star-forming cores, disks and outflows, and supernova remnants.

On Galactic size scales (i.e. covering the full Galactic Plane), the only studies to date have been carried out by the Planck satellite, which mapped polarised dust emission at 345\,GHz on large angular scales over the entire sky (see e.g., \citealt{planck2015_intXIX})
suggesting a high polarisation fraction (up to 15--20\%). On the scales traced by Planck at 5$'$ resolution, the polarisation fraction systematically decreases with column density, but this could be largely due to the effects of small-scale field tangling. This clearly indicates how the low angular resolution of the Planck data limits the study of the Galactic magnetic field, because it can only be traced at scales of 0.2 to 0.5\,pc for even the closest clouds, as shown in Figure~\ref{fig:cra}.  This points to the need for high-resolution (i.e. a few arcseconds) observations both to resolve the complex magnetic fields within star-forming clouds and to investigate the evolution of dust grain properties in high-density regions.  High-angular resolution sub-millimeter observations of star-forming regions such as those carried out with the SMA and ALMA have obtained a very detailed picture of the role of magnetic fields from filaments to disks and outflows, but 
they have been limited only to a few representative clouds or small samples of cores (see, e.g., \citealt{hull2013}; \citealt{zhang2014}; \citealt{liu2020}),
while single-dish observations have been restricted to high-surface-brightness areas within molecular clouds: the JCMT BISTRO Survey \citep{wardthompson2017},
the largest existing polarimetric survey of star-forming clouds, has mapped an area of  $\sim$1.5\,deg$^2$ at 14$^{\prime\prime}$ resolution. Therefore, to gain a statistical understanding of the role of magnetic fields in star formation and the evolution of the ISM, and to reach conclusive results on their importance, origin and evolution, it is necessary to map polarised emission continuously from Galactic scales through to individual dense star-forming cores at high angular resolution. For this, the proposed capabilities of AtLAST would be both unprecedented and essential. 

\subsection{Technical Justification}

We propose an unbiased survey with AtLAST of the whole visible Galactic plane in dust polarised thermal emission, covering $-210^\circ<l<60^\circ $, with $\left | b \right |< 1^\circ$ at 230 GHz, 345 GHz, and 460 GHz. Such a survey will provide, for the first time, high-angular resolution observations of Galactic dust polarisation over a large area. AtLAST is the only facility able to carry out a systematic large-scale survey of magnetic fields that directly connects cloud, filament and core fields thanks to its mapping speed and angular resolution. In fact, the angular resolution provided by AtLAST (6.5$^{\prime\prime}$--3.3$^{\prime\prime}$ at 230--460\,GHz) will allow us to study the polarised emission from cloud scales to core scales (0.1 pc) across the Galactic Plane to the distance of the Galactic Center. For the closest clouds, the angular resolution of AtLAST will allow us to trace the magnetic field well within the embedded cores (e.g., $\sim$0.02--0.03\,pc (4000--6000 au), depending on the wavelength, for clouds at 1\,kpc).

This large-area coverage survey will trace the morphology and direction of magnetic field lines, and the degree of polarisation in the molecular ISM from cloud to core scales, which will allow us to constrain, in a robust statistical way, the role of magnetic fields in cloud evolution and gravitational fragmentation as a function of both cloud mass and star-forming environment of our Galaxy. In particular, we will unveil the structure of the magnetic field at the different spatial scales and particle density levels, following the agglomeration process of the matter from the diffuse medium of the molecular clouds to their denser filamentary network populated by dense clumps and cores where stars form. The unprecedented angular resolution of AtLAST will allow us to easily complement the interferometric studies of selected regions at smaller scales. The survey will use multi-chroic continuum cameras and will observe the polarised emission of the Galactic Plane at three different wavelengths, 1.3\,mm, 0.87\,mm, and 0.65 mm, which correspond to 230\,GHz, 345\,GHz, and 460\,GHz, respectively. This will allow us to study the polarisation spectrum, which is a powerful probe of dust grain properties \citep{hensley2021},
and to  estimate the spectral index of the emission, which will allow us to separate the free-free contribution from the dust continuum emission in high-mass star-forming regions, and to unveil important information regarding the efficiency of alignment and/or possibly grain-growth \citep{santos2019}.

A key, and unique, feature of this survey will be its unbiased, high-resolution continuum observations of the Galactic plane.  Single-dish observations of magnetic fields in star-forming regions have hitherto largely been restricted to nearby, typically low-mass, clouds ($d<0.5$\,kpc for the best-resolved and best-studied clouds, where $d$ is heliocentric distance), due to current limitations on resolution and mapping speed (e.g. \citealt{pattle2017}; \citealt{santos2019}).  However, the majority of stars do not form in clouds such as these, instead preferentially forming in high-mass, cluster-forming clouds which, being less common than low-mass clouds, are typically more distant and so preferentially closer in projection to the Galactic plane (e.g. \citealt{motte2018}).  Observations of magnetic fields in more distant high-mass star-forming regions with ALMA have an extremely small field of view (e.g. \citealt{fernandezlopez2021}; \citealt{sanhueza2021}), and do not allow us to understand the role of magnetic fields of directing gas flows onto star-forming hubs.  By observing at high resolution with AtLAST in the Galactic Plane, we will be able to map magnetic fields in the hearts of the cluster-forming massive clouds where the majority of stars form, and the connections of these fields to the large-scale magnetic field in the surrounding molecular ISM.  We will further be able to investigate the magnetic fields in regions within the molecular ISM that have been affected by feedback from massive star formation, such as supernova shocks and expanding H\textsc{ii} regions.
 
We plan to detect polarisation fractions as low as 2\%, which is the typical polarisation fraction measured by \textit{Planck} in the highest-density regions of molecular clouds (\citealt{planck2015_intXIX}), across the Galactic plane.  It should be noted that polarisation fraction typically decreases with increasing gas density (\citealt{whittet2008}; \citealt{jones2015}; \citealt{pattle2019}), from a maximum of $\sim 20$\% in low-density molecular gas to $\sim 2$\% in the densest regions \citep{planck2015_intXIX}, so by detecting polarisation in the densest gas, we ensure a detection throughout the molecular ISM.  We aim to detect polarised emission for sources with dust emission peak fluxes in total intensity as low as 100\,mJy\,beam$^{-1}$, which corresponds to a peak flux density in polarised intensity of 2\,mJy\,beam$^{-1}$ (a brightness temperature of 1.1\,mK), with a signal-to-noise of 3 on the 2\% polarised intensity. Assuming an emitting area of 0.1\,pc, the typical core size, and different distances, we have used the current version of the AtLAST sensitivity calculator to estimate the integration times for different core distances and at the different wavelengths. For our calculations we consider a dish diameter of 50\,meters, a total bandwidth of 32\,GHz, and a surveyed area of 540\,deg$^2$.  %

\vspace{50pt}
\section{The Dynamics of Our Galaxy}
\label{sec:Dyn}

\newcommand{\Dyn}[1]{\DTLfetch{projects}{Parameters}{#1}{Galactic_Dynamics}}

\subsection{Scientific Rationale}
With advancing surveys of the Galactic Plane that detail the gas kinematics at different scales \cite[for a compliation see][]{Stanke2019}, such as single-dish surveys of the inner Galactic disc \cite[e.g. SEDIGISM on APEX,][]{Schuller17} and dense clumps \cite[e.g. MALT90 on Mopra,][]{Jackson13}, a paradigm is emerging: the star-formation process can be modelled as a continuous flow of mass and energy at all scales, with no physical boundaries among different structures such as clouds, filaments, clumps and cores. This flow mediates star-formation thanks to a continuous interplay between gravity, turbulence and magnetic fields. This mutual interplay can substantially change with the environment, which complicates our ultimate goal of building a comprehensive description of the star-formation mechanism, in particular for high-mass stars. The dynamics of the flow seems indeed to differ between extremely different regions of our own Galaxy: peculiar behaviours are seen in unique environments such as the Central Molecular Zone \citep{Henshaw19}, and indications of different mechanisms of the star formation process seem to arise when we compare results from e.g. the inner versus the outer Galaxy \citep{Olmi23}, or between in arm and inter arm regions \citep{Eden13}. But still much more statistically relevant work has to be done to understand these first indications. 

The shape of the initial mass function (IMF) of stars implies that although high-mass stars are rare, when they do form there is a significant population of lower-mass stars formed in the same region.  This means that the most common environment in which stars like the Sun form is a clustered environment. In fact, radioactive isotope products in meteorites have indicated that the Solar system was likely born in a cluster with high mass stars (Adams 2010, Dauphas \& Chaussidon 2011).  Thus, the nearby low-mass star forming regions which we can study in detail are in fact the exception, and not the rule. To form a general understanding of star formation requires capturing the environments and conditions under which the plurality of stars form. The large average distances to these regions have hampered studies to date.  Not only do high-mass star forming regions harbour significant numbers of low-mass objects at various stages of formation, but the stars with masses above 8 M$_{\odot}$ themselves play a fundamental role in this paradigm by injecting energy and feedback back into the medium which then regulates and shapes the Milky Way.   
From the theoretical point of view, some studies suggest that the massive star-formation process is a so-called ``core-fed'' scenario  \cite[e.g.][]{McKee03}, while others suggest a ``clump-fed'' scenario which is much more dynamical \citep{Anderson21}. These streams are regulated mostly, if not completely, by the gravitational potential from one side \citep{Vazquez-Semadeni19} or by the turbulent eddies from the other \citep{Padoan20}. Observationally, there are examples of regions following both scenarios \citep[e.g.][]{Henshaw14,Peretto13}, but overall it seems that the role of gravity does indeed increase significantly as we look at denser and denser regions \citep{Ballesteros-Paredes11,Anderson21}, with the densest ones dominated by ordered inflow motions from filaments down to cores \citep{Traficante20}. Surely, the multi-scale flow seems to be the key to understand the final formation of stars in a given region \citep{Montillaud19}.

The above shows that even in our Galaxy we are not yet able to systematically (or in a statistically significant way) probe the most common mode of star formation: in clustered environments which often include the effects of nearby massive stars. %
By virtue of sampling more of the initial mass function, high-mass star-forming regions are the regions in which most low-mass stars form, and thus, most sun like stars formed under the influence of feedback from their nearby high-mass neighbours.  We need to understand these `typical' star forming environments as a population to be able to understand how our Solar System came to be. In these regions, where energy feedback can often be an order of magnitude greater than in nearby low-mass star-forming regions (e.g. Reiter et al. 2020), it is unclear how the hotter, more energetic environments affect the next generation of stars.

\begin{figure*}%
    \centering
    \includegraphics[width=1\textwidth]{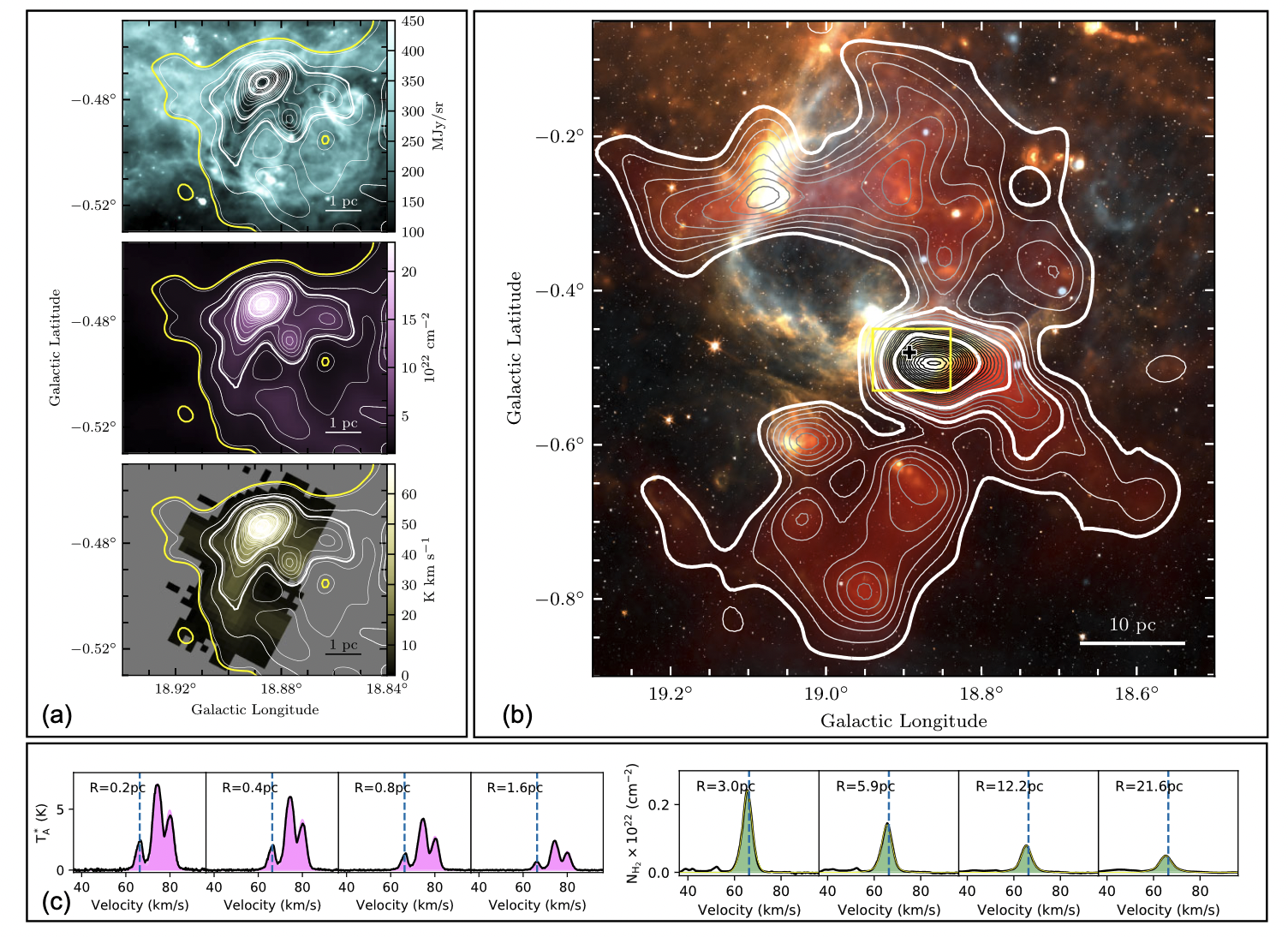} %
    \caption{ Images of SDC18.888-0.476. (a) Top – Spitzer 8 $\mu$m; middle – H$_{2}$ column density from Herschel observations; bottom – N$_{2}$H$^{+}$ (1-0) integrated emission. The contours are identical in all panels, and are those of the H$_{2}$ column density image. The yellow contour corresponds to edges of the N$_{2}$H$^{+}$ (1-0) emission. The four thicker white contours are those used to compute the average N$_{2}$H$^{+}$ (1-0) spectra displayed in magenta in panel. (c), (b) Multicolour image of the molecular cloud hosting the SDC18.888-0.476 infrared dark clump (white: 3.6 $\mu$m, orange: 8 $\mu$m, yellow: 70 $\mu$m, orange: 350 $\mu$m, blue: 1.42 GHz, red: H$_{2}$ column density). The contours show the H$_{2}$ column density obtained from the Galactic Ring Survey $^{13}$CO(1–0) data. The thicker white contours are those used to compute the $^{13}$CO(1–0)-based spectra shown in green in panel (c). The plus symbol shows the central position of the IRDC, and the yellow rectangle shows the coverage of the images displayed in (a). (c) Spectra averaged within the highlighted H$_{2}$ column density contours in panels (a) and (b). The radius of the region within which the spectra have been averaged are indicated in each panel. The vertical blue dashed lines show the systematic clump velocity as measured from N$_{2}$H$^{+}$ (1-0). From \citet{Peretto23}.}%
    \label{fig:Peretto_23}
\end{figure*}

All these studies are based on a few selected regions in the sky, but to properly understand the details of the interplay between gravity and turbulence at the various spatial scales and gas densities one needs to produce Galactic Plane surveys which covers tens of parsecs and with a resolution down to the sub-parsec scales, covering a wide range of physical environments, and able to trace both the physical properties and the gas dynamics of each and every star-forming region in the Galaxy.

\subsubsection{State of the art and the role of AtLAST}
The physical properties of large and intermediate scales can be derived from observations in the far-infrared (FIR)/sub-mm
regime with surveys such as the Herschel Galactic Plane survey Hi-GAL \citep{Molinari2010}. This project has revealed $\simeq30,000$ filaments across the Plane \citep{Schisano20} and more than $\simeq150,000$ clumps which will likely form the new generation of stars \citep{Elia17,Elia21}.  These continuum based observations cannot however, uncover the dynamics of these regions: an understanding of how the gas is moving. 

The dynamic signatures must be derived from the emission of specific molecular lines able to trace the various density structures. At the \textit{tens} of parsec scales, various CO surveys such as SEDIGISM \citep{Schuller17}, or CHIMPS/2 \citep{Rigby19,Eden20} are now providing insights in the gas dynamics in a statistically significant way for the parts of the surveyed regions where the CO is still optically thin (where densities are of order 10$^3$ - 10$^4$ cm$^{-3}$). On the \textit{tenths} of parsec scales, a survey of the densest clouds in our Galaxy (ALMA-IMF) \citep{motte2022} and the largest survey of star-forming regions carried out with ALMA (ALMAGAL, Molinari et al., in prep.), are now providing unique information of the physical and dynamical properties of the innermost cores. What is still missing are 1) a large survey to link these scales by studying the dynamics at the intermediate ones, which requires mapping dense gas tracers able to penetrate into the clouds at the clump scales, and 2) Galactic scale surveys linking those targeted single dish surveys (e.g. SEDIGISM and CHIMPS/2) of parts of the plane to their full Galactic context.  A large single dish facility with a large field of view can achieve both of those aims at the same time. 

To advance in this field, we envisage an AtLAST survey that maps, for the first time, the whole visible Galactic Plane in the atmospheric window equivalent to ALMA Band 3 (i.e. at 3mm), with the aim of tracing the gas dynamics with an unprecedented combination of spatial and spectral resolution, sensitivity and angular coverage. This survey will observe tens-of-parsecs long clouds and filaments down to the clumps and sub-clumps structures up to 10 kpc away from us covering a wide range of temperatures and column densities. These molecules are able to penetrate the clouds to higher densities than the CO observed in previous studies. Such a survey will provide the first comprehensive and statistically significant study of the gas dynamics in star-forming regions in different Galactic environments (arm vs. interarm regions, inner vs. outer Galaxy, etc).  The combination of these observations with the dust continuum observations presented in Section \ref{sec:Pol} will allow to fully characterize physical quantities such as virial parameter and mass accretion rate and will allow for testing different gravo-turbulent scenarios.

\subsection{Technical Justification}
We propose different configurations for this survey which allow us to fulfill increasingly significant scientific goals. The most comprehensive survey aims to map the visible portion of the Galactic Plane $-210^{\circ} < l < 60^{\circ}$, with $\vert b\vert < 1^{\circ}$ at $\simeq90$ GHz ($\simeq3$ mm). The main lines of this project are optically thin ones such as N$_{2}$H$^{+}$ (1-0) ($\simeq 93.3$) GHz and H$^{13}$CO$^{+}$ (1-0) ($\simeq86.7$ GHz), and optically thick ones such as HCO$^{+}$(1 - 0) ($\simeq 89.2$ GHz) or HNC(1 - 0) ($\simeq90.7$ GHz). In addition, the wide bandwidth provided by AtLAST 
will allow us to add other tracers in a single tuning. They will complement the study of the local dynamics in these regions, such as investigating the impact and the energetics of the feedback from massive stars through shock tracers such as SiO (2-1) ($\simeq86.8$ GHz). For our survey we require a spectral resolution of 0.1 km/s, which will allow us to resolve the hyperfine components of N$_{2}$H$^{+}$ (1-0) and the double peaks in blue/red-shifted profiles of optically thick tracers such as HCO$^{+}$(1 - 0) or HNC (1 - 0).

Thanks to the capabilities of AtLAST we aim to reach a sensitivity of $\simeq0.03$K per channel, equivalent to $\simeq45$ mJy assuming the beam size of a 50m dish. Such a sensitivity per channel is sufficient to trace the emission of the main lines of this project in each star-forming region of the Galactic Plane, based on previous observations of N$_{2}$H$^{+}$ (1-0) and HCO$^{+}$(1 - 0) lines \citep[e.g.][]{Traficante17}. 

Finally, the spatial resolution of the AtLAST 50m dish will allow us to reach an unprecedented angular resolution at $\simeq3$mm of $\simeq15$", opening up the opportunity to look at sub-parsec structures up to 10 kpc away from us ($\simeq0.7$ pc resolution at 10 kpc). At the same time, we will trace sub-core scales in the nearby regions, with a resolution of $\simeq0.01$ pc at a distance of 2 kpc. 

Assuming a survey area of 540 deg$^{2}$, the total on source integration time for this Galactic Plane survey is $\simeq3400$ hours according to the AtLAST sensitivity calculator, as shown in Table \ref{tab:galactic_dynamics}. This calculation considers a receiver with 1000 pixels, an average source elevation of 25$^{\circ}$ and a system temperature of T$_{sys}=60$ K. The full survey in combination with ancillary data will allow us to reach a deep understanding of the Galactic dynamics in different environments: we will fully cover the outer galaxy (quadrants III and IV covering 180$^{\circ}$ < $\ell$ < 360$^\circ$), 30$^{\circ}$ of galactic quadrant II and up to 60$^{\circ}$ of galactic quadrant I, including the Central Molecular Zone. The full survey will allow the study of the dynamics as function of spiral arms, extreme environments (e.g. the tip of the bar, or the Galactic Center) and inner vs. outer Galaxy. Alternatively, a survey area of 360 deg$^{2}$ will include the longitude range $-120^{\circ} < l < 60^{\circ}$. Such a survey will still allow us to reach most of the main scientific goals, at the expense of the comparison between the dynamics in the inner and in the outer Galaxy and it will require a total of $\simeq2300$ hours. Finally, we consider the possibility of a much more limited survey, covering only 120 deg$^{2}$ which will cover the inner part of the Galaxy, including the tip of the bar and the Galactic Center, at the cost of $\simeq750$ hours of observing time.

\begin{table*}[hbt]
\centering
\caption*{Galactic dynamics integration times}
\begin{tabular}{ccc|cc}
\hline
Area  &Longitude&Velocity& Sensitivity & Integration   \\
coverage & coverage& Resolution & & time \\
(deg$^{2}$)&($^{\circ}$)&(km s$^{-1}$)&(K) & (h) \\
\hline 
540 & -210$^{\circ}\leq$\textit{l}$\leq60^{\circ}$ & 0.1 & 0.03 & 3400  \\
360 & -120$^{\circ}\leq$\textit{l}$\leq60^{\circ}$ & 0.1 & 0.03 & 2300  \\
120 & $\vert$\textit{l}$\vert\leq60^{\circ}$ & 0.1 & 0.03 & 750  \\

\end{tabular}
\caption{On source integration times for different areas of the Galactic Plane covered by the proposed Galactic dynamics survey at the sensitivities and velocity resolutions stated in the column headings.}
\label{tab:galactic_dynamics}
\end{table*}

\section{The Chemistry of Our Galaxy}
\label{sec:Eco}
\newcommand{\Eco}[1]{\DTLfetch{projects}{Parameters}{#1}{Galactic_Ecology}}
\newcommand{\BALG}[1]{\textcolor{blue}{#1}}

\subsection{Scientific Rationale}

Over the years, we have made progress in the study of isolated low-mass star-formation, learned more about clustered star-formation, and refined our knowledge about evolved stars, but, as outlined below, we are missing a sense of how those processes are connected to their larger scale environments. This includes scales from fractions to tens of parsec (core to Giant Molecular Cloud, GMC) scales.  Key diagnostics of these connections are coupled to the chemistry and dynamics of the gas in the Galactic Plane and the properties of the accompanying dust population.  It is this lifecycle of gas and dust that shapes the evolution of Galaxies, and ours is no exception. Deep, whole cloud studies of the chemistry in these regions sets the stage for the detailed analysis of the star forming core size scale.

An AtLAST survey aimed at better understanding the lifecycle of gas and dust needs to quantify the key chemical constituents of that gas to understand the dominant heating and cooling processes which then help shape the evolution in this complex system. The rotational transitions of molecules like CO and other diffuse gas tracers discussed below give vital information about how material becomes available in cores for star and planet formation, but it is only when paired with chemical surveys at higher frequencies that we unlock the ability to observe key heating and cooling lines (i.e. [CI] and CN in diffuse gas), diagnostics of higher temperatures (i.e. H$_2$CO) and CO Spectral Line Energy Distributions (SLEDs).

The (sub-)mm atmospheric windows provide a unique opportunity to trace gas from diffuse to intermediate to dense regimes. In particular, in both the mm and sub-mm ranges there are networks of  simple species and multiple transitions of each. Of particular importance are the [CI] lines at 492 and 809 GHz which can be used to trace the CO-dark gas in molecular cloud envelopes \cite{Clark12,Wolfire10}. Further, there are numerous lines from CN throughout each band, but of particular utility are the lower-energy lines at 113 GHz. AtLAST also has broad access to many CO lines from $^{12}$CO, $^{13}$CO and C$^{18}$O in their  J = (1-0) to (9-8) transitions. The higher-J lines are of particular importance since they are sensitive to excitation in energetic environments, such as shocks, photon dominated regions (PDRs), X-ray dominated regions (XDRs) and  cosmic-ray dominated regions (CRDRs) \cite[see e.g.][]{Wolfire22,Gaches22}. Finally, with access to high frequency observations, simple hydrides ions can also be detected, such as ArH$^+$ (at 617 GHz) which can also trace diffuse CO-dark gas \citep{Schilke14}. %
Access to many lines of molecular tracers of different scales with the same facility will enable AtLAST to pursue science goals inaccessible to other facilities.

 Here, using the 230 to 690 GHz windows, we aim to probe the low- to high- density gas using CO (in multiple transitions from J= 2-1 to 8-7), and high density gas with species such as HCO$^+$, CS, and their isotopologues.  H$_2$CO (formaldehyde) is an excellent temperature tracer \citep{mangumwootten1993} in moderate to high density regions and therefore highly complimentary to the N$_2$H$^+$ thermometer available at lower (86 GHz) frequencies. Shocks can be traced by SiO (J = 8-7 and 5-4), and other less prominent complex organic molecules as and when they fall into our spectral bandwidth.  In addition to these species, which trace densities from 10$^2$ cm$^{-3}$ to 10$^8$ cm$^{-3}$, we will also be able to simultaneously detect a number of their isotopologues (i.e. $^{13}$CO, C$^{18}$O, H$^{13}$CO$^+$, and C$^{34}$S) which tend to be significantly less abundant (by orders of magnitude) and therefore do not become optically thick until much higher densities for a given transition --- allowing us to probe the abundance, chemistry and dynamics of a system at higher densities and smaller scales.  With large ($\sim$32 GHz) bandwidths, many of these tracers can be observed simultaneously.  For instance, in the tuning with CO J = 4-3, assuming a 32 GHz bandwidth means that  CS J = 10-9, C$^{34}$S J = 10-9, HDO and CI can be observed simultaneously by tuning between $\sim$460 and 492 GHz.

The fine structure lines of [CI] at 492 \& 809 GHz are important tracers of molecular gas, initially thought to be predominantly from Photo Dissociation Regions (PDRs) but later shown to be well-mixed with molecular hydrogen and CO \citep[e.g.][]{little94,glover2015,glover2016}. [CI] is a particularly important tracer of lower density gas \citep[$\le 5 \times 10^{4}$ cm$^{-3}$,][]{glover2016}, PDRs, and cosmic ray ionisation. Moreover, [CI] has distinct advantages over CO in tracing low metallicity gas \citep{glover2016} where its relationship to H$_{2}$ density is linear and constant. Due to its favourable redshifted frequency range and the difficulty of observing CO at low metallicities [CI] is a popular tracer of gas in high redshift galaxies \citep{papadopoulos2004}. 

Unfortunately, despite the utility of these lines, their observations have been restricted to small scale studies at either very low angular resolution ($\sim$arcmin) or of limited samples at higher resolution \citep[e.g.][]{lee2022,beuther2020}. Hence, we have limited understanding of the total carbon budget across the Milky Way ISM, especially of the $\sim$30\% of molecular gas that is not traced by CO \citep[the so-called CO-dark gas,][]{seifried2020}, of low metallicity gas, or of gas irradiated by high cosmic ray intensities \cite[see, for example][]{Bisbas15,Bisbas17,Bialy15}. This is particularly problematic not only for understanding the molecular ecology of our own Galaxy but also for understanding the physical processes at work in galaxies in the more distant Universe. Observations of [CI] across a much wider range of Galactic environments are crucial to understanding the carbon ecology of the ISM.

With the combined power of the spatial resolution, large FoV and high spectral resolution of AtLAST, we can probe the chemistry of the ISM in a statistical way across our Galaxy, targetting specific Giant Molecular Clouds and their surroundings,  delving into the questions below:

\begin{itemize}
    \item Quantifying the chemistry of the interstellar medium (ISM) gives insights into the dominant heating and cooling processes in various systems, which then links to their total energy budgets. 
    \item Quantifying the bulk chemistry of (unresolved) protostellar and protoplanetary disks in nearby regions gives insights both into the thermal history of the star/planet formation process as well as an indication of what types of volatiles and complex organic molecules are available in these systems. \item Deriving the heating/cooling budget of the regions being observed using CO SLEDs and the other tracers mentioned above
\end{itemize}

With a large, systematic survey, not only will we be able to probe these atomic and molecular species, but with a full Galactic plane survey we will be able to see how their abundances vary as a function of position in the Galaxy in a statistical way. Metallicity enhancements or reductions affect how a system cools as it collapses, since emission lines are some of the primary cooling mechanisms in regions like star forming cores.  There are indications of metallicity gradients in the Galaxy \citep[e.g.]{Milam05,Bolatto13}, but they are based on targeted surveys to known star forming regions. An unbiased survey gives us the best chance of understanding the morphology of any gradients which is the first step towards understanding its nature.

In addition to the chemistry, with high spectral resolution we can also quantify the dynamics of the gas in giant molecular clouds as described in Section \ref{sec:Dyn} above. Both the chemistry and dynamics give insights into what the future of a given system will be.  How is the bulk of the material moving? Is it static? Is there expansion or contraction? Rotation, infall or outflow? The dynamics of the various gas components (be they ionised, atomic or molecular) tells us about what has happened, is happening, and what will happen in a system. Infall or contraction motions are often identified by double peaked profiles in optically thick lines (with single peaks in the optically thin transitions at the rest velocity of the source), rotation is, on small scales, traced by butterfly diagrams, and on larger scales by shifts in velocity fields.  Outflow or expansion is generally traced by line wing emission.  Being able to disambiguate these types of motions requires high (sub-km/s) spectral resolution. On the larger scales, this includes the motion of the gas along or within filaments (see Section \ref{sec:Dyn}). On intermediate scales, the rotation of a collapsing protostellar core gives an indication of the mass enclosed, and the expansion velocity of a supernova remnant or HII region gives an indication of the energy required to push against the ambient material. On small scales, those probed by observatories like ALMA, we reach the size scales of protostellar and protoplanetary disks to look at the details of how planetary systems form.

Pulling information together on all of these processes and size scales will only be possible with a facility like AtLAST. This gives us our first statistical insights on the ecology of our own Galaxy: how environment and interactions shapes the next stages of its evolution.

It is not only the birth of stars that shapes the evolution of our galaxy, but their lives and deaths as well. AGB stars and supernovae are responsible for dust evolution and enrichment of the ISM. That, combined with the explosive energy releases at the end of their lives, leaves fertile ground for the next generation to begin. Further, winds and, finally, deaths of high-mass stars populate the ISM with both stable and radioactive material, the latter of which, in particular $^{26}$Al, are important for the current and next generation of planet formation models \citep{2019Lichtenberg,2021Diehl}.

The chemistry, dynamics, and recycling of material all tie directly to the gas, dust and magnetic properties of the ISM and allow us to quantify the energy transfer between those populations, their shielding from external (or internal) photoionising radiation, the interplay between gravity and magnetic fields in these regions, and what the likely next stages will be in their evolution. By studying the \textit{ecology} and evolution of the Milky Way, we look back at our origins in the aim of understanding our future.

With AtLAST, we can statistically quantify evolution of material from being cold and diffuse (as traced by CI) through to warm (as traced by CO and Complex Organic Molecules, COMs), star forming, ionized (as traced by Radio Recombination Lines), chemically rich (as traced by COMs), and shocked gas which feeds back onto the environment in the forms of PDRs and the like, setting the stage for the next generation of star formation and energy transfer back into the broader ISM. Current surveys cannot reach the CO dark, cold gas that makes up the bulk of the diffuse ISM, nor can they capture COMs in a statistically significant number of regions. Getting sufficient sensitivity and coverage of the Galactic Plane in these ecological surveys,  with a complete census of the multi-phase ISM, enables a detailed understanding of the evolution of our galaxy; not only as a template for others.

\begin{table*}[hbt]
\centering
\caption*{Galactic Chemistry `wedding cake' integration times}
\begin{tabular}{ccc|ccc|ccc}
&&& \multicolumn{3}{c|}{Full Plane (-210$^\circ < \ell < 60^\circ$, $b=\pm$0.6)} & \multicolumn{3}{c}{Square Degree}\\
&&& \multicolumn{3}{c|}{(0.3 K, 0.1 km/s)} & \multicolumn{3}{c}{0.03 K, 0.01 km/s}\\
\hline
Central &ALMA&AtLAST& Single & Pointings & Full & Single &Pointings& Full \\
Freq. & Band& Resolution &Pointing & Required &Area & Pointing & Required &Area\\
(GHz)& &($''$)&($\mathrm{s}$) & &($\mathrm{h}$) & ($\mathrm{m}$) && ($\mathrm{h}$) \\
\hline 
115.3 & 3 & 13.1 & 3.4 & 24752 & 23.4 & 56.7 & 77 & 72.2 \\
230.5 & 6 & 6.5 & 1.1 & 98919 & 29.9 & 18.2 & 306 & 92.4 \\
345.8 & 7 & 4.4 & 2.0 & 222619 & 121.2 & 32.7 & 688 & 374.1 \\
461.0 & 8 & 3.3 & 7.7 & 395743 & 844.1 & 128.0 & 1222 & 2605.2 \\
691.5 & 9 & 2.2 & 23.5 & 890190 & 5816.4 & 392.0 & 2748 & 17952.0 \\
\end{tabular}
\caption{On source integration times for single tuning, full plane (shallow)  and 1 sq. deg (deep) surveys at the sensitivities and spectral resolutions stated in the column headings at the frequencies of the CO J=1-0, 2-1, 3-2, 4-3, and 6-5 transitions. The number of pointings required to image the full areas described are based on the assumption of a 1000 pixel spectroscopic instrument. Applying a `wedding cake' approach enables shallow surveys of the whole plane with follow-ups in the regions worth deeper integrations to derive the chemistry. For reference, in recent observing cycles, ALMA has planned for $\sim$4000 hr of observing per year.}
\label{tab:chemistry}
\end{table*}

\subsubsection{State of the art and the role of AtLAST}

From a single dish point of view, the only molecular survey to cover the majority of the Galactic Plane is the CfA 1.2 m telescope survey of  CO \citep{Dame2001}. With 7.5$'$ resolution, it was the first set of surveys to show the overall intensity of CO gas in the Galactic Plane as well as giving the first position-velocity (PV) diagram of the bulk gas in the ISM. No higher-resolution full-plane follow-ups have been possible, making it extremely challenging to build a coherent picture of what the ISM is doing across the Galaxy.

There are heterogeneous attempts to study parts of the Galactic Plane, making the most of current capabilitites \cite[for an overview of their coverage, see Figure 1 of][]{Stanke2019}. For example, surveys such as SEDIGISM \citep{Schuller17,Schuller2021} have started to look at the bulk gas of the Galactic Plane by focusing on $^{13}$CO, for a 84 deg$^{2}$ portion of the inner Galaxy.  Because it is less optically thick than the more common $^{12}$CO isotopologue, it is a better tracer of the cold and dense ISM observable with the resolution limits of the APEX telescope (28$''$ at 220 GHz).  The sensitivity of this survey (T$_{mb}$ = 0.8-1.0K at a spectral resolution of 0.25 km s$^{-1}$) means that a number of less abundant species that fall within the $\sim$4 GHz spectral bandwidth of these observations could also be observed, although, with the exception of C$^{18}$O,  they are only detected in the brightest regions. These species often trace different gas components to the $^{13}$CO, and include CH$_3$CN\footnote{A dense gas tracer often used with ALMA to study the kinematics of the disks around O and B type forming stars\citep{Cesaroni17}}, SiO\footnote{A tracer of low velocity shocks \citep{Schilke97}}, and other complex organic molecules \citep[see Table 1 in ][]{Schuller17} with the kinds of low excitation energies that could be expected to emit on the size scales probed by APEX.
FYST (formerly CCAT-p) will be conducting an ecology study of its own, focusing on [CI], CO(4-3) and higher frequency transitions in a 200 sq. deg field at a spectral resolution of 0.5 km s$^{-1}$ and sensitivity of 0.25 K (1.0 km s$^{-1}$ and 0.1 K in the Magellanic clouds). %

From an interferometric point of view, there are detailed studies like those of PILS and EMoCA \citep[][respectively]{PILS,emoca} that have studied the chemistry of individual objects in great detail.  From these works we can get a sense of what sensitivity limits are required in order to detect the chemical signatures of star forming regions in the larger AtLAST beams. In these examples, these highly sensitive chemical surveys were able to achieve 0.1-0.2 K rms noise levels, which we adopt here as what is necessary to quantify the complex chemistry of the ISM and star forming regions.  

One of the key science drivers of the ALMA 2030 upgrade program is to detect and characterise the chemistry in Galactic star forming regions. We know where these regions are in a global sense from surveys like HIGAL, GLIMPSE, ATLASGAL etc, but from these continuum surveys at low spatial resolution, we still cannot pinpoint the most interesting (and/or most significant for our understanding of galactic evolution) pointings for ALMA follow-up. When building follow-up interferometric surveys, one of the key elements to success is knowing which targets to observe. Without large scale, highly sensitive surveys of the Galactic plane, seeding the next generation of high-resolution, targeted follow-up surveys becomes more and more challenging.

\subsection{Technical Justification}

\subsubsection{Sensitivity requirements}
What is currently not possible is to combine the wide area surveys like SEDIGISM to the sensitivities required to quantify the complex heating, cooling, and overall chemistry of star forming regions at a spatial resolution sufficient to identify individual cores across most of the Galaxy.  The high spectral resolution ($ \leq 0.1 \, \rm km\, s^{-1}$) is essential to spectrally resolve the lines. In the cold and quiescent phases of the ISM ($T \lesssim 20\,$K) that AtLAST will resolve, molecular lines are as narrow as $0.3\, \rm km\, s^{-1}$.  Resolving them with at least 3-5 channels is crucial to both perform high-sensitivity kinematics studies and to recover the line fluxes (and, hence, molecular column densities) correctly.

To push the limits of large-scale Galactic Plane surveys to those of the deeper ALMA studies and reach a 1 $\sigma$ limit of 0.03 K ($\sim $20 mJy) in a 0.01 km s$^{-1}$ channel would take 18 min of on source integration for a single field at 230 GHz (ALMA Band 6) and would allow us to detect species such as c-H$3$H$_2$, H$_2$CO, CH$_3$OH and HC$_3$N which can be quite extended around cores and their surrounding filaments. Such studies are bound to revolutionise our current knowledge of the chemical budget available in the early stages of star-formation. Large maps will enable the study of the chemical differentiation of these species allowing to explore the interconnection of chemical and physical structures in a large range of scales. Assuming the spectral line receiver on the telescope will have 1000 pixels, at this frequency we would require about 90 hr of on source time per square degree assuming we can observe all of the important species in a single tuning, which necessitates wide bandwidth receivers. Table \ref{tab:chemistry} shows how that scales through observing all of the rotational transitions of CO from J=1-0 to J=6-5 to build up the CO SLED, while simultaneously observing the other species mentioned above due to the large (16 GHz+) anticipated bandwidths of the spectral line receivers. 

\subsubsection{"Wedding Cake" approach to chemical surveys}

Even with a 50m class single-dish facility with a 1000 pixel spectral line camera, to complete a deep (0.03 K) spectroscopic survey of the entire plane is not possible. However, completing a shallower but (spatially) complete survey which already significantly improve on current best surveys is possible.  This necessitates the use of a `wedding cake' approach to chemical surveys, where shallower, larger area surveys are used to derive where to drill down into the regions that show a clear need for follow-up, deeper observations.  Table \ref{tab:chemistry} shows these extremes of the shallow (although still twice as deep as SEDIGISM) and deep limits, where the shallower survey (1$\sigma$ = 0.3 K) has broader (0.1 km s$^{-1}$) channels, and the deeper (1$\sigma$ = 0.03 K) has narrower (0.01 km s$^{-1}$) channels. This  shows that drilling down into the chemistry of our galaxy  by an order of magnitude in both depth and spectral resolution will be feasible for regions of the sky identified in the shallower surveys.  With an order of magnitude greater number of pixels on the spectral line instrument, the numbers in Table \ref{tab:chemistry} also drop by an order of magnitude, making deep, full plane, multi-band observations possible.

\section{Planetary System Evolution}
\label{sec:discs}

Stars form from cores and begin their main-sequence lives surrounded by a protoplanetary disc of gas and dust. 
Over a few million years, planetesimals form from the dust and, in some parts of the discs, go on to form planets \citep{Williams+2011, ALMA2015, Andrews2018, Drazkowska2023}. 
Remnant dust and gas is either accreted by the star/planets or dispersed from the disc, e.g., via processes such as radiation pressure or disc winds \citep[see e.g.,][]{Ercolano2017, Owen2019, Pascucci2023}. 
As the gas dissipates, the mutual gravitational interaction of planetesimals and any planetary companions dynamically stirs the planetesimals, initiating a collisional cascade resulting in the liberation of dusty material visible in scattered light and/or thermal emission, and potentially molecular gas \citep{Wyatt2008, Marino2022}. 
This dust is a rich laboratory revealing the present state and past evolution of the host planetary system, and is commonly referred to as a debris disc.

Debris discs are commonly viewed as extra--Solar analogues to the Kuiper belt in the Solar System \citep{Kuiper1951, Fernandez1980, 1984Aumann}.
Much like the physical connections made between Solar System planetary orbital dynamics and the structure of the Kuiper belt \citep[see e.g.,][]{Fernandez1984}, circumstellar debris disc emission can be used to infer the exoplanetary system histories.
Indeed, to study the structure of debris discs, and thus their connection to the broader planetary systems of which they are one such component, requires an understanding of the orbital dynamics of their collisional dust. 
Although debris discs can be detected at optical and infrared wavelengths, observations at those wavelengths are primarily sensitive to small (micron--sized) dust grains that are susceptible to transport forces.
Moreover, at these wavelengths, the stellar light dominates that of the debris disc. 
Submillimetre observations provide the best opportunity for studying large dust grains that are still co-located with their parent planetesimals (and thus the orbits of planetesimals in Kuiper belt analogues), and at a wavelength whereby a debris disc's emission dominates that of its host star.

Submillimetre observations provided us with some of the first resolved images of debris discs \citep{1998holland, Wilner2002, Greaves2005, 2017Holland}. 
In particular, observations with the ALMA interferometer have revolutionised our understanding of debris discs (and indeed discs at earlier evolutionary stages) by providing the deepest millimetre maps of circumstellar dust which have unlocked clues as to the evolution of extra--Solar planetary systems \citep[see e.g.,][]{2014Dent, 2017Macgregor, 2018MarinoGap, Lovell2021c, 2023Booth}.

In this section we explore how transformational advances in our understanding of planetary systems necessitate a new large aperture single-dish facility. 

\begin{itemize}
    \item Furthering our understanding of the epoch of planet and planetesimal formation requires large surveys of young stellar objects (YSOs) in star forming regions, the efficiency of such surveys substantially benefiting from the combination of a high sensitivity and large field of view, and the higher frequencies in the sub--millimetre domain, closer to the peak emission wavelengths of cold (10s of K) dusty discs. 
    \item Furthering our understanding of mature debris discs as faint as our own Kuiper belt in order to place our Solar System into context requires surveying the nearest stars at high sensitivity and with a resolution of a few arcseconds, whilst retaining sensitivity to structures of tens of arcseconds. 

\end{itemize}
Here we will demonstrate how a new facility like AtLAST can address both these scientific domains.

\subsection{Dispersal of primordial discs and the birth of debris discs}
\subsubsection{State of the art and the role of AtLAST}
Typical nearby ($\leq$ 300 pc) star-forming regions have 100s--1000s of young stellar objects \citep[YSOs; which are dominated by `class~III' YSOs, i.e., the most evolved young stellar objects, that typically have dispersed the bulk of their primordial dust and gas, see e.g.,][]{Ortiz-Leon+2018, Luhman2020, Krolikowski+2021, Luhman2022}.
Class~III YSOs are at an evolutionary stage when the bulk of their giant planets, terrestrial planet--embryos, and planetesimals (extra--Solar analogues to kilometre--sized comets and asteroids) have formed \citep{Wyatt2015}.
As such, these YSOs bridge the evolutionary period over which planetary systems go from systems in formation through to mature, evolved systems around stars on the stellar main--sequence that host planets and debris discs.
Class~III YSOs therefore offer a view into the early stages of development of young planetary systems, with observations at millimetre wavelengths being sensitive to their cool (10s of K) circumstellar dust components during this transition, unlocking views on their typical masses, morphologies, grain--size distributions, and dispersal mechanisms \citep{Lovell2021b}.

However, the number of published constraints on the presence of millimeter emission from class~III stars is poor, with $
{<}100$  having been observed with ALMA \citep{Michel+2021}, and with only small-number samples from other millimeter instruments having been surveyed (such as recent class~III surveys with the Submillimeter Array (SMA) at 1.3\,mm; e.g., Lovell et al. (in prep), Andrews et al., (in prep) whereby only 10s of class~III YSOs have been observed at a time, due to either reduced sensitivity and/or slow mapping capabilities. 
Despite their promise for understanding the early evolution of planetary systems, and the discovery space for the youngest populations of debris discs, class~III stars remain severely under-studied at (sub-)mm wavelengths, primarily driven by a lack of  (sub-)mm observatories with sufficient mapping speeds to observe 100s--1000s of objects on short cadences, with a sensitivity to the faint emission from their typical dust levels.

For nearby stars within a few 10s of pc, millimeter imaging (with ALMA) can now readily obtain dust--mass sensitivities down to $0.001\,M_\oplus$ \citep[see e.g.,][]{Wyatt2008, Matthews+2014, Hughes+2018}, and $0.02-0.1\,M_\oplus$ for nearby star--forming regions out to 150\,pc \citep[see e.g. the Lupus surveys of][]{Ansdell+2016, Ansdell+2018} on relatively short timescales (observations spanning a few to a few tens of minutes).
Indeed, for ground-based instruments, (sub-)mm wavelength observations are the best tracer of bulk cold material in discs (i.e., the dust and gas reservoirs at 10s--100s of au), which (when optically thin) provide constraints on a disc's total mass budget \citep[see e.g.,][]{Hildebrand1983, AndrewsWilliams2005, Ansdell+2016, Lovell+2021a}.
Importantly, only moderate spatial resolution data (e.g. $\lesssim 4-6''$) are required in order to derive disc masses, due to i) the availability of optical and near-/mid-infrared all-sky data from Gaia, 2MASS and WISE (which in conjunction can help to uniquely tie any observed emission to the location of a given observed class~III star), and ii) the typical separations of YSOs being much greater than this angular distance.
Furthermore, studying discs across multiple frequencies within the sub-mm bands enables studies of their grain--size distributions, and thus in young developing systems, an understanding of whether these can be better understood as `remnant' primordial discs, or young debris discs; a key question at their earliest stage of development.
Without a thorough understanding of the typical masses of discs at the class~III YSO stage, %
we are missing crucial information as to how early planetary systems assemble in their first few, to first few tens of Myr. 
Measuring and characterising a large sample of discs with their disc mass distributions and multi--band (sub--)millimeter emission profiles will open a brand new window on the formation and evolution of exo-planetary systems.%

\subsubsection{The role of AtLAST in detecting the evolution from protoplanetary to debris discs}
AtLAST, with its large field-of-view (FoV), large format continuum cameras and 50\,m aperture will enable a step-change in our understanding of the formation of debris discs around class~III stars, and their eventual evolution towards the stellar main sequence. 
With a 2$^{\circ}$ FoV (filling 10s of thousands of ALMA primary beam areas), and simultaneous deep, multi--band continuum coverage (to comparable and/or better sensitivity on the same per--pointing timescales as ALMA), AtLAST offers a unique opportunity to study this aspect of planetary system evolution.

\begin{figure*}[htb]
    \centering
    \includegraphics[width=\textwidth, trim={4cm 0cm 0cm 0cm},clip]{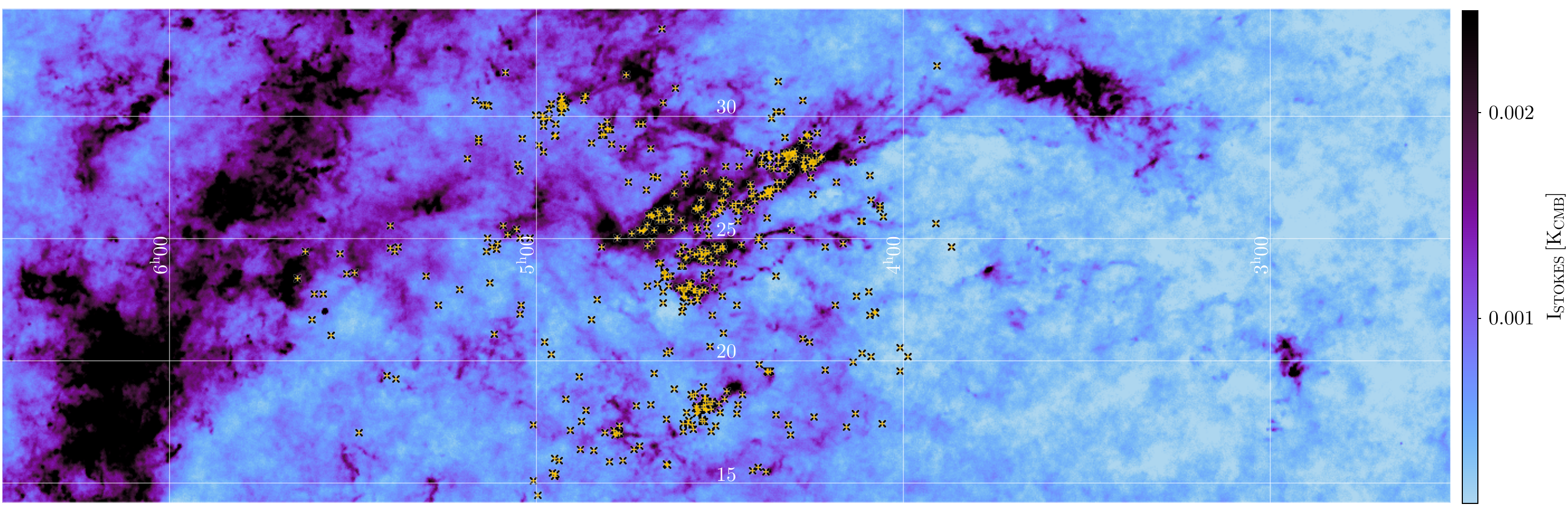}
    \caption{A thermal dust map \citep[sourced from the IRSA database][]{PlanckPR1} of Planck (PR1), and over-plotted, all YSOs in the Taurus star-forming region identified by \citet{Krolikowski+2021} as yellow--black cross-hairs. As seen, there are many separate regions with densely populated YSOs, i.e., that within a $2\times2$--degree field dozens are located. Note the rectangles in this plot show $15\times5$--degree boxes (each of which are covered by around 19 AtLAST pointings).}
        \label{fig:taurusYSOs}
\end{figure*}

With its large FoV, and ability to instantaneously map significant portions of nearby star-forming regions in one pointing, AtLAST will enable the complete mapping these regions on timescales hitherto unachievable at (sub-)mm wavelengths; factors of 20-500 times shorter than ALMA with its current capabilities, and 5-120 times shorter than ALMA with its anticipated upgrades \citep[see, e.g.,][]{Carpenter+2023}. These numbers were derived using the AtLAST sensitivity calculator and the known or expected ALMA sensitivities, a fiduciary star forming region size (i.e. Taurus, see Figure \ref{fig:taurusYSOs}), and the ranges reflect the speedups from longest wavelengths ($\sim$1mm) to shortest ($\sim$ 0.35mm).

Recognising the significant potential for AtLAST to study early planeteary system evolution, we propose a survey of the Sun's nearest planet-forming neighbours, mapping  planet-forming potential of all ${<}300\,$pc YSOs, to sensitivities enabling detections of young planet--forming discs in multiple (sub-)mm bands.
We stress that such a survey is currently beyond the technical ability of any existing instruments.
From the catalog of \citet{Kerr+2023}, we note that there are 33 star-forming regions (SFRs) south of $+40^\circ$-decl, within 1\,kpc.
To map these 33 SFRs in full, would require approximately 1000\,hours of on source time with AtLAST at all four specified frequencies, completing deep observations of tens of thousands of class~III YSOs.
Because of the large telescope throughput (high sensitivity across a large field of view), these observations would also enable mapping of earlier stage YSOs in the same fields as well as measure their millimetre variability \citep[the subject of much recent attention, see e.g.,][]{Johnstone+2022, VargasGonzalez+2023, Lovell2024} if the observing cadence / repeats are properly spaced. Variability studies more generally are presented separately in Orlowski-Scherer et al. (in prep.).

\subsection{Debris discs in nearby planetary systems}
\subsubsection{State of the art and the role of AtLAST}
The first debris discs were detected with the \textit{InfraRed Astronomical Satellite}, identifying infrared excesses for a handful of objects \citep[e.g.][]{1984Aumann,1993BackmanParesce}. 
Further advances with the \textit{Infrared Satellite Observatory} and \textit{Spitzer} gave insights into the evolution of debris discs \citep[e.g.][]{2001Habing,2006Su,2007Wyatt,2011Kains,2011Moor}. %
Although infrared excesses can enable estimates of dust temperature and radii, any such derivations are riddled with degeneracies, due to poorly understood quantities, such as the optical properties of the dust, and dust size distributions, which lead to large errors in the true size and temperature properties of discs \citep{Pawellek+2015}.

Early resolved observations demonstrated the variety of disc structures with features such as warps \citep{kalas95}, eccentricities \citep{kalas05} and clumps \citep{1998Greaves}, hinting at influence from unseen planets. 
With \textit{Herschel}, we were able to systematically resolve 10s of debris discs, greatly improving our understanding of the optical properties of the discs \citep[e.g.][]{2013Booth,2014Pawellek,2016Morales,2021Marshall}. 

These infrared observations are dominated by small grains that are susceptible to transport forces that push them away from the star or drag them towards it. 
To investigate the parent planetesimal population requires observations at sub-mm or longer wavelengths. 
The JCMT Legacy Survey SONS \citep[SCUBA-2 Observations of Nearby Stars][]{2017Holland} surveyed the 100 brightest debris discs, detecting half of them and thereby doubling the number of discs detected in the (sub-)mm. 
Amongst other results, these observations provided constraints on the (sub-)mm slope of their spectral energy distributions, which can in turn be used to determine the size distribution of grains \citep[e.g.][]{2020Lohne}.
However, the resolution and sensitivity of the JCMT means that SONS was only able to follow up known debris discs, not create an unbiased survey. 
This is particularly true for discs around low mass M dwarf stars. 
These stars make up the majority of stars in our galaxy and yet very few discs have been detected around them \citep{2009Lestrade,2018Kennedy,2020Luppe, 2023Cronin}.

\begin{figure*}
    \centering
    \includegraphics[width=0.9\textwidth]{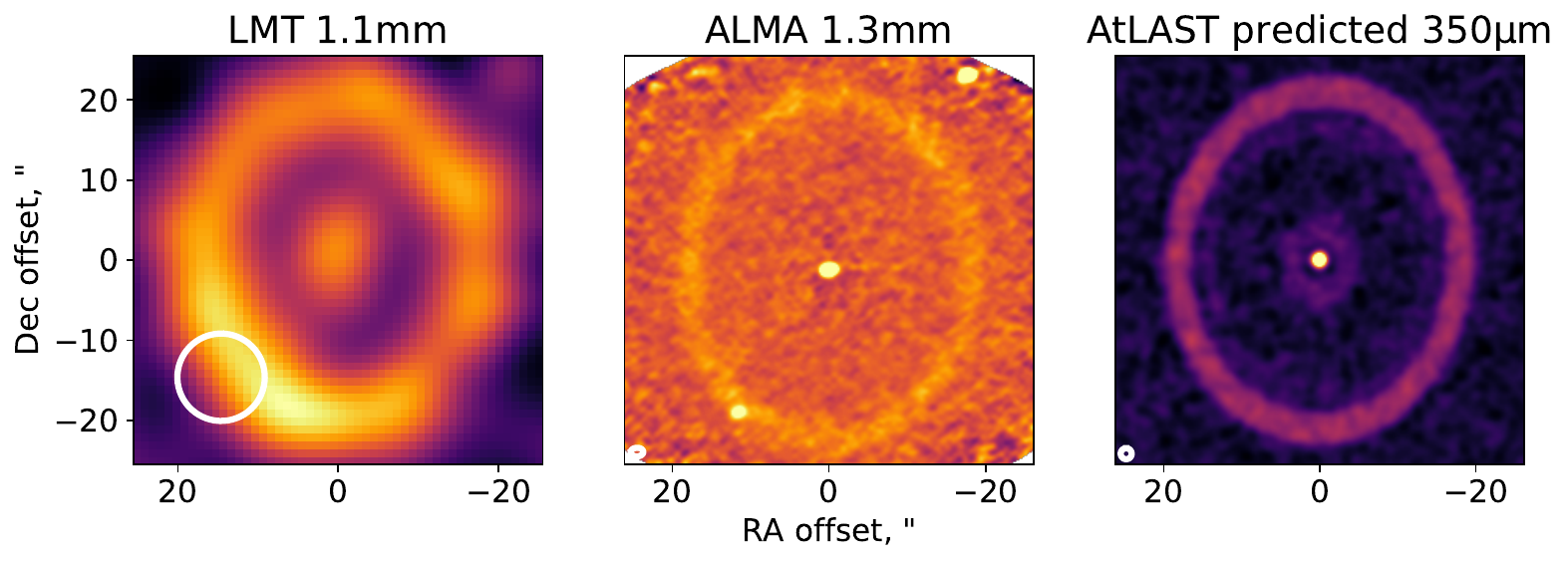}
    \caption{Observations of $\epsilon$ Eridani with LMT \citep{2016Chavez} and ALMA \citep{2023Booth} along with a predicted image using AtLAST for observation with an on source time equivalent to the ALMA image. For a nearby disc like this, AtLAST can reach scales of a few au much more rapidly and without loss of large-scale flux that hinders such observations with ALMA.
    }
    \label{fig:epEri}
\end{figure*}

Further follow-up observations with ALMA have revolutionised our understanding of debris discs through its capability to produce high spatial resolution images of many bright and nearby debris disc systems \citep[see][for a recent review]{Hughes+2018}.
However, interferometry as a technique has two weaknesses; the high resolution dilutes the disc flux over many beams frustrating detection of the faintest discs around nearby stars, and as an interferometer it lacks sensitivity to large scale structure. 
These limitations are illustrated by observations of the closest known debris disc, $\epsilon$ Eridani, shown in Figure \ref{fig:epEri}.
With ALMA the star is clearly visible, but the dust emission is suppressed as the high resolution results in a low SNR and the disc size (40 arcseconds) is larger than the maximum recoverable scale of ALMA at this wavelength (12 arcseconds) whereas both the star and dust are recovered (albeit at lower s/n and resolution) with LMT observations at comparable wavelengths \citep{2016Chavez,2023Booth}.

Over the past 40 years, the field of debris discs has expanded from simple photometry to detailed characterisation of select systems. These are invariably the brightest systems, many orders of magnitude more massive than the Kuiper belt. What the field needs now is a large survey of nearby stars in the sub-mm reaching levels as low as the Kuiper belt with multi-wavelength observations with a resolution of a few arcseconds to find and characterise this population of faint discs for comparison with the known, but not neccessarily typical brighter discs. With a large, single dish sub-mm telescope we will be able to simultaneously achieve the high resolution and sensitivity to large-scale structure necessary for this \citep[see also][]{2019Holland}.

\begin{figure}
    \centering
    \includegraphics[width=0.45\textwidth]{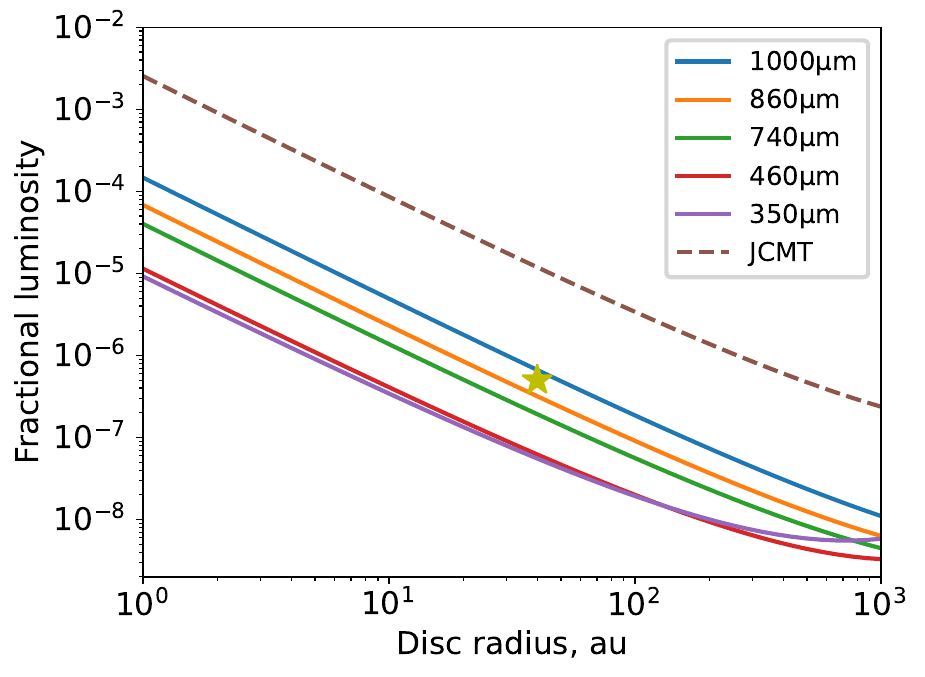}
    \caption{Fractional luminosity limits vs. radius comparing AtLAST with JCMT for a G2V star at 10~pc. Confusion limits \citep[calculated from][]{2017Bethermin} are shown for AtLAST wavelengths 740, 860 and 1000$\mu$m. For wavelengths 350 and 460$\mu$m, the confusion limit is too low to be defined and so we instead assume a 20 hour observation to demonstrate how deep these observations can go. For JCMT (at 850$\mu$m), the line corresponds to the confusion limit from \citet{2017Geach}. 
    The level of Solar System's Kuiper belt is shown by the yellow star based on the results of \citet{2019Poppe}. 
    }
    \label{fig:kuiperbelt}
\end{figure}

\subsubsection{The role of AtLAST in detecting and characterising planetary systems}

To make transformational advancements in the field of debris discs the main considerations are sensitivity to faint sources, ability to resolve the discs and multi-wavelength observations within the (sub-)mm atmospheric windows.

First considering the sensitivity requirements, a common observational measure of the dust in a debris disc is the fractional luminosity. This is the ratio between the bolometric luminosity of the disc and that of the star. The Kuiper belt is estimated to have a fractional luminosity of $5\times10^{-7}$ based on results of 
the New Horizons mission \citep{2019Poppe}, whereas debris discs that have been detected so far are typically at least an order of magnitude brighter \citep{2016Montesinos,2018Sibthorpe} and those detected at sub-mm wavelengths specifically are at least two orders of magnitude brighter \citep{2017Holland}. %
 The main limitation on reaching such low fractional luminosities with a sub-millimetre single-dish telescope is the extragalactic confusion limit \citep[confusion with sources within our own galaxy is also possible, but is found to be negligible at these wavelengths][]{2002Blain}. 
 Here we are considering the photometric confusion limit, which we take to be 3 times the confusion noise. The confusion noise is calculated from the simulations of \citet{2017Bethermin}\footnote{see the white paper on the distant universe, van Kampen et al., in prep.}. %
 This limit on the flux density can be converted to a fractional luminosity limit using equation 8 from \citet{Wyatt2008}. In figure \ref{fig:kuiperbelt}, we show this for AtLAST compared to JCMT and LMT. Using the AtLAST sensitivity calculator\footnote{\url{https://github.com/ukatc/AtLAST_sensitivity_calculator} assuming 25\% weather profile, dual polarisation and broad bandwidths utilising the full atmospheric window as described in Di Mascolo et al., in prep.}, we find that we reach these limits in a matter of minutes for observations >1mm and hours at 740 and 860$\mu$m, whilst at the shortest wavelengths, the confusion limit is undefined. This demonstrates not only the need to reach low flux density levels but also the major benefit of sub-millimetre observations with a 50m telescope in our quest to observe discs that can be considered true Kuiper Belt analogues.

ALMA's high resolution means that the impact of confusion noise is limited, and the sensitivity of the main array approximately matches that of AtLAST, which might initially give the impression that it is more suitable to finding the faintest discs. Unfortunately, the higher resolution (4 times higher than AtLAST in the compact configuration) means spreading the emission over multiple beams and weakening the signal. In addition, as an interferometer, it is limited in its ability to capture faint, extended emission. This is quantified by the maximum recoverable scale (MRS), which ranges from 3 to 8 arcsecs for the main array in compact configuration. Meaning that any discs with a $r>d*\rm{MRS}/2$ will be missing flux and this effect can be problematic at shorter scales too \citep{2023Plunkett}. This demonstrates the major benefit of a single-dish telescope for finding faint sources.

With a resolution of 4.2 arcseconds at 860~$\mu$m, we will be able to resolve discs as small as the Kuiper belt around stars within about 20~pc. Additional observations at 350~$\mu$m will allow us to reach scales as small as 1.7 arcseconds (a few au around the nearest stars), providing a large sample of disc radii - greatly expanding on the results of the ALMA REASONS survey (Matr\`a et al., in prep.). Combining multi-wavelength observations will also provide us with accurate constraints on the spectral energy distribution. When combined with constraints on the disc size from these observations and disc composition from optical and infra-red observations, this can be used to determine the size distribution. 

The advances in detection rates that will be possible with AtLAST make an unbiased survey of the nearest stars in the submillimetre viable for the first time. 
For instance, we can survey the nearest 500 stars (100 each for spectral types A, F, G, K and M) along the lines of the survey proposed in \citet{2007Matthews}. 
In terms of detecting faint discs, 860~$\mu$m observations will allow us to reach Kuiper belt levels of dust around Sun-like stars within an hour of observing time per star (going down to a sensitivity of 21~$\mu$Jy\,beam$^{-1}$ -- 3 times the confusion noise). 
For M dwarfs, this is expected to increase the number of detected discs by an order of magnitude (Luppe, in prep.). 
Additional observations at 350~$\mu$m are necessary to reach spatial scales as small as 1.7 arcseconds (a few au around the nearest stars), thereby resolving all discs as small as the Kuiper belt around stars within about 20~pc. 
Combining multi-wavelength observations also provides a measure of the spectral slope necessary to our understanding of collisional processes within debris discs. 
This survey will greatly increase our knowledge of planetary systems in the solar neighbourhood and provide us with the statistics necessary to determine how disc properties correlate with stellar properties. In addition, these observations will also be useful for studying the stars themselves as the multi-wavelength observations enable the study of chromospheric and coronal contributions to the atmosphere \citep{liseau13,mohan21} as well as the possibility of picking up stellar flares \citep{macgregor20}. This is explored in more detail in a companion paper by Wedemeyer et al. (in prep.).

\section{Summary of telescope requirements}
\label{sec:telescope}

Together, the science cases presented above show a clear need for new facilities to understand the galactic ecosystem in which we live.  Along the way, they have identified key requirements an observatory would require in order to fulfill the scientific need.  Below we break down these technical requirements for such a future (sub-)mm single dish facility from the point of view of Galactic astronomy.

\subsection{Need for a large FoV}
The Milky Way stretches across the entire sky. Naturally, this motivates a large FoV, both when mapping the entire Galactic plane or smaller regions such as star-forming regions. The bulk of the Galactic Plane is observable in the Southern Hemisphere, and with an observatory sharing the same sky coverage as ALMA, we can survey from -210$^\circ$ < $\ell$ < 60 $^\circ$.  Most of the material in the Galactic Plane is within 1$^\circ$ of the midplane, which means that a Galactic Plane survey would need to cover \textit{up to} 540\,deg$^2$.  The larger the FoV, and the larger the cameras (both spectral and continuum), the fewer individual pointings are required to survey the entire plane (See, for instance Table \ref{tab:chemistry}) or individual nearby star forming regions (as shown in Figure \ref{fig:taurusYSOs}), where the spatial extent is of order a few square degrees.

\subsection{Need for a single-dish telescope}
The large scale mapping described above necessitate the use of single dish facilities because interferometers suffer from a lack of sensitivity to large-scale emission, typically quantified by the maximum recoverable scale. For ALMA in its compact (low resolution) configurations at sub-mm wavelengths, this scale is of order a few arcseconds. This means that the large scale structures we anticipate studying would be completely resolved out by interferometers. For example, the nearest known extrasolar debris disc has a diameter of around 40 arcseconds, still well above the recoverable scales for the ALMA main array alone (see Section \ref{sec:discs}). For targets on these scales, a single-dish telescope is necessary to ensure an accurate measure of the flux and avoid any biases introduced by interferometric artifacts.

\subsection{Need for a 50m diameter}
A 50m diameter telescope observing in the sub-mm provides two key benefits over modern facilities; resolution and sensitivity. With this diameter, AtLAST can achieve sub 10$''$ resolution at all sub-mm wavelengths. From a technical point of view, this brings down confusion limits significantly from current sub-mm observatories, and with 4 beams across the width of the ALMA instantaneous field of view, followup interferometric observations can be properly targeted.  From a scientific point of view, this allows for the disambiguation of star forming cores from their environments out to 10 kpc (i.e. further than the Central Molecular Zone of the Galactic Centre). %

In order to detect and study the faintest structures in the galaxy, the high sensitivity made possible by a 50m diameter single dish sub-mm telescope is necessary, because collecting area has a direct relation to achievable sensitivity. Reaching line sensitivities of order 0.1 K means we can probe gas temperatures (using, for example N$_2$H$^+$ fine structure lines, see Section \ref{sec:Dyn}) across most star forming cores. To detect debris discs as faint as the Kuiper belt around the nearest stars requires reaching an RMS on the order of 10s of $\mu$Jy/beam at sub-mm wavelengths, which is below the confusion limit of current sub-mm facilities.

\subsection{Need for sub-mm capabilities}

As described above, there are a lot of rotational transitions of important molecules that emit at sub-mm wavelengths; some tracing the thermal balance of the gas, others tracing its ionisation state or the dynamics of the region under study.  These lines, when combined with observations of some of their mm counterparts (i.e. full plane low-J CO surveys, see Section \ref{sec:Dyn}) give insights into the history, current state, and future of the gas in our Galaxy - telling us about how this ecosystem is evolving and shaping the next generation of stars.

For the denser components of the Giant Molecular Clouds in our Galaxy, sub-mm observations of the dust continuum are critical for capturing the shape of the spectral energy distribution (SED), and, as shown in Section \ref{sec:Pol}, are far less likely to be contaminated by the synchrotron emission coming from ionised regions that can contaminate (or dominate) the mm SED.

\subsection{Instrumentation requirements}

\subsubsection{Continuum camera}
The Galactic polarisation, disc evolution and nearby stars cases all require broadband continuum observations in order to reach the deep sensitivities to dust emission, with polarisation preserving instrumentation.  A multi-chroic camera would be advantageous for observing in the Galactic Plane because the size scales being probed are small enough that time variability within the observed regions could be a large source of uncertainty in any calibration (see a companion paper by Orlowski-Scherer et al., in prep.). This would come at the cost of overall mapping efficiency, as the number of pixels available for any one waveband would necessarily decrease.

\subsubsection{Spectral line observations}

Key to understanding the chemistry and dynamics of our Galaxy, and the star forming regions within it, is observations of large numbers of spectral lines. Large format (i.e. 1000 pixel) heterodyne instruments take advantage of the large field of view described above, but in order to efficiently detect the species described in Sections \ref{sec:Dyn} and \ref{sec:Eco}, we need high spectral resolution (down to 0.01 km s$^{-1}$) spectrometers with large bandwidths. A 16 GHz simultaneous bandwidth means that moderate and high density tracers like CO (3-2) and HCO+ (4-3) can be observed simultaneously in Band 7 - something ALMA cannot currently do.  With a 32 GHz band, CO (6-5) can be simultaneously observed with [CI], a known tracer of 'CO-dark' molecular gas.  Simultaneous observations of these types of species reduces systematics in the calibration, and minimises uncertainties due to variability in these regions.

\subsection{Considerations for Telescope Operations and Data Reduction}

The variety of science cases described here require very different types of observational setup; from large scale, largely automated, surveys of the sky to dedicated principal investigator projects lasting only a few hours.  Scheduling the telescope must accommodate both in order for these science cases to progress.

The large format continuum and spectral line cameras above will create significant volumes of data, especially if the spectral line cameras can achieve the 0.01-0.1 km s$^{-1}$ spectral resolution required for Galactic observations across a 32 GHz bandwidth. AtLAST will not be alone with this problem.  With ALMA's Wideband Sensitivity Upgrade \citep{Carpenter+2023}, and the even larger fields of view possible with the SKAO, data storage, reduction and analysis will become significantly more challenging as instrumentation improves.  

\section{ Summary and derived telescope requirements}
\label{sec:summary}

Above, we have summarised the state of the art in terms of what is possible to know about our Galaxy from a 'whole system' point of view; we are limited by the sensitivity, resolution and mapping capabilities of current facilities. To make progress in understanding how the Galaxy around us is evolving requires a new telescope with orders of magnitude increases in capabilities.

The chemical and dynamical surveys of our Galaxy aim to understand the interplay between gravity and turbulence at a number of key scales covering from protostellar cores up to giant molecular clouds. The Galactic polarisation project aims to expand this into understanding the roles of magnetic fields and dust continuum in those equations.  The YSOs and debris discs studies then focus in on detecting planetary systems and understanding their evolution - possibly even detecting, for the first time - the equivalent of the Kupier belt in other nearby systems.

Table \ref{tab:matrix} provides a summary of the telescope and instrument requirements presented above in Section \ref{sec:telescope}. None of the science cases presented here can be done with current generation facilities, but they are vital for understanding the evolution of our Galaxy. 

Overall, the ambition of these projects far exceeds the capabilities of current and near-term facilities. No currently operating, or under construction facility can address these questions about our origins and place in the dynamic Universe.  A 50m class sub-mm single dish telescope with a large field of view populated with mega-pixel continuum (polarisation) cameras possible of simultaneous multi-chroic observations and kilo-pixel high-spectral resolution (e.g. 0.1 km s$^{-1}$ or higher) instruments all of which have bandwidth comparable to those expected in the ALMA Wideband Sensitivity Upgrade (e.g. 32 GHz) will enable us to engage with the next generation questions raised above (about dust, magnetic fields, chemistry and dynamics), and address future challenges that remain to be revealed.

With a telescope like AtLAST we can determine the overall, yet detailed, ecology of our own Galaxy. This includes determining the kinematics and intensity of the magnetic fields from large to small star-forming scales while at the same time deriving the dust properties on the same scales. Probing the dynamics of that gas and dust then adds the 6D structure of the materials: where they are (3D position), and what they're doing (3D momentum) to statistically determine what will happen next. With these types of surveys we lay the foundation for where to dig more deeply into the chemistry of the Galaxy.

Taking an overall view of our Galaxy, we can answer a number of interesting questions: Is chemistry the same across the Galaxy, and how do changes in chemistry affect cooling?  What are planetary discs made of across the Galaxy, and are they consistent? At what stages do pre-biotic materials start to appear? How is the dense gas moving in relation to its environment?   We can then add to that the ability to resolve many distance ambiguities through kinematic distance analyses, and breaking the temperature/density degeneracies in continuum surveys by being able to take the temperature of the gas we probe.

\begin{table*}
\begin{tabular}{p{2cm}|p{1.5cm}|p{2cm}|p{1.9cm}|p{2cm}|p{2cm}|p{2cm}}%
\hline
\bfseries Parameters & \bfseries Units & \bfseries Polarisation & \bfseries Dynamics & \bfseries Chemistry & \bfseries YSOs & \bfseries Nearby Discs%
\csvreader[head to column names]{projects.csv}{}%
{\\\hline \csvcoli & \csvcolii & \csvcoliv & \csvcoliii & \csvcolv & \csvcolvii & \csvcolvi} %
\end{tabular}
\caption{Telescope parameters required by the science topics presented in this paper. when presenting properties for the various topics, the most stringent were used in this table (i.e. the deepest dynamics and chemical studies).}
\label{tab:matrix}
\end{table*}

\DTLloaddb{projects}{projects.csv}

\newcommand{\unit}[1]{\DTLfetch{projects}{Parameters}{#1}{Units}}

\section*{Data and software availability}

The calculations used to derive integration times for this paper were done using the AtLAST sensitivity calculator, a deliverable of Horizon 2020 research project `Towards AtLAST', and available from \href{https://github.com/ukatc/AtLAST_sensitivity_calculator}{this link}.

\section*{Grant information}
This project has received funding from the European Union’s Horizon 2020 research and innovation programme under grant agreement No. 951815 (AtLAST).
K.P. is a Royal Society University Research Fellow, supported by grant number URF\textbackslash R1\textbackslash 211322.
S.M. is supported by a Royal Society University Research Fellowship (URF-R1-221669)
J.B.L. acknowledges the Smithsonian Institute for funding via a Submillimeter Array (SMA) Fellowship. 
L.D.M. acknowledges support by the French government, through the UCA\textsuperscript{J.E.D.I.} Investments in the Future project managed by the National Research Agency (ANR) with the reference number ANR-15-IDEX-01.
M.L. acknowledges support from the European Union’s Horizon Europe research and innovation programme under the Marie Sk\l odowska-Curie grant agreement No 101107795.
S.W. acknowledges support by the Research Council of Norway through the EMISSA project (project number 286853) and the Centres of Excellence scheme, project number 262622 (``Rosseland Centre for Solar Physics'').

\begingroup
\small%
\bibliographystyle{apj_mod}
\setlength{\parskip}{0pt}
\setlength{\bibsep}{0pt}
\bibliography{bibliography}
\endgroup

\end{document}